\documentclass[twocolumn]{aastex631}

\usepackage{color}
\usepackage{hyperref}
\usepackage[shortlabels]{enumitem}

\shorttitle{1D binary properties}
\shortauthors{Pourmand and Ivanova}

\graphicspath{{./}{figures/}}

\begin{document}

\title{Properties of binary systems in a one-dimensional approximation}

\author[0000-0002-5414-243X]{Ali Pourmand}
\affiliation{Department of Physics, University of Alberta, 
Edmonton, T6G 2E7, Alberta, Canada}

\author[0000-0001-6251-5315]{Natalia Ivanova}

\affiliation{Department of Physics, University of Alberta, 
Edmonton, T6G 2E7, Alberta, Canada}

\correspondingauthor{Natalia Ivanova}
\email{nata.ivanova@ualberta.ca}

\begin{abstract}
Evolutionary calculations for stars in close binary systems are in high demand to obtain better constraints on gravitational wave source progenitors, understand transient events from stellar interactions, and more.
Modern one-dimensional stellar codes make use of the Roche lobe radius $R_{\rm L}$ concept in order to treat stars in binary systems. If the stellar companion is approaching its $R_{\rm L}$, mass transfer treatment is initiated.  
However, the effective acceleration also affects the evolution of a star in a close binary system. This is different from the gravity inside a single star, whether that single star is rotating or not. Here, we present numerically obtained tables of properties of stars in a binary system as a function of the effective potential:  volume-equivalent radii of the equipotential surfaces, effective accelerations and the inverse  effective accelerations averaged over the same equipotential surfaces,  and the properties of the $L_1$ plane cross-sections.
The tables are obtained for binaries where the ratios of the primary star mass to the companion star mass are from $10^{-6}$ to $10^5$ and include equipotential surfaces up to the star's outer Lagrangian point.
We describe the numerical methods used to obtain these quantities and report how we verified the numerical results.  
We also describe and verify the method to obtain the effective acceleration for non-point mass distributions. We supply a sample code showing how to use our tables to get the average effective accelerations in one-dimensional stellar codes.
\end{abstract}


\section{Introduction}
Binary stars are stellar systems consisting of two stars that are gravitationally bound together and orbiting around each other.
If the radius of one of the stars during the course of its evolution becomes comparable to the orbital separation, the binary is termed a close binary.
In a close binary such phenomena as tidal spin-up, stable mass transfer (MT), or unstable MT (common envelope evolution) can occur. The evolution of each star in a binary system can be significantly altered from the evolution of a similar but evolved-in-isolation star. 
Substantial attention is now given in modern one-dimensional (1D) stellar codes to treat the evolution of stars on their way toward the start of MT. The structure
of the donor star, specifically of its envelope, at the start of MT -- when the volume of the donor star is approaching the volume of its Roche lobe -- plays a crucial role in determining whether the MT proceeds stably or unstably.

Upon approaching contact, each star in a binary system is affected strongly by the gravitational field of its companion and by the binary's orbital motion.
However, the primary effect that is considered by 1D stellar evolutionary codes is the size of the Roche lobe of the donor star. Recently, it has become appreciated that the donor star may remain in a state of substantial Roche lobe overflow (RLOF)\footnote{As substantial RLOF, we mean an overflow by 10-20 percent of the donor Roche lobe radius, where the donor is approaching the effective potential surface passing through by the donor's outer Lagrangian point.} for an evolutionary-noticeable time while keeping MT in a stable regime, appearing, for example, as Ultra-Luminous X-ray sources \citep{Pavl2017}. Furthermore, the binding energy of the donor's envelope at the onset of CE, following the MT while the donor significantly overfills its Roche lobe,  plays an essential role in the initial conditions of three-dimensional (3D) simulations of common envelope events \citep{2020cee..book.....I}. When a star is close to its Roche lobe radius, or overfills it, the outer parts of the stellar envelope are strongly affected by the effective acceleration in a binary system, and this effective acceleration is different from the one that a single star experiences. While the knowledge of what is happening to the star in this regime has been around for about 60 years\citep{kopal1959close}, this physics has not yet been included in detail in 1D stellar calculations.

The main goal of this paper is to provide the community with the database we have constructed. That database contains the various properties of binary stellar systems and the code that allows anyone to easily use  binary physics in their future binary or MT numerical studies using 1D codes. 
We review in \S~2 the physics that becomes relevant when simulating the gravitational field of a binary star. In \S~3 we discuss the assumptions made to approximate binary stars into a 1D scheme, and the numerical methods employed to obtain volume-equivalent radii and average gravitational accelerations for our tables; we also obtain some analytical expressions which determine how this gravitational acceleration behaves close to the center of the donor star, and allow verification of the numerical method in this regime. In \S~4 we verify our results by comparing them to various published results and present self-convergence checks. In \S~5 we discuss the effects of a non-point mass on the binary potentials, and how to use our tables in the case of non-point masses. 
This Paper I is devoted to the numerical methods only.
The application of the tables that we obtain for binary evolution, and the scientific outcomes, are described in the follow-up Paper II.

\section{Binary in a corotating frame}

Here we review the theory of binary's effective potential and introduce the specific quantities we obtain numerically and present in the database.

\subsection{Effective binary potential.}

\begin{figure*}[p]
  \centering
\includegraphics[width=\columnwidth]{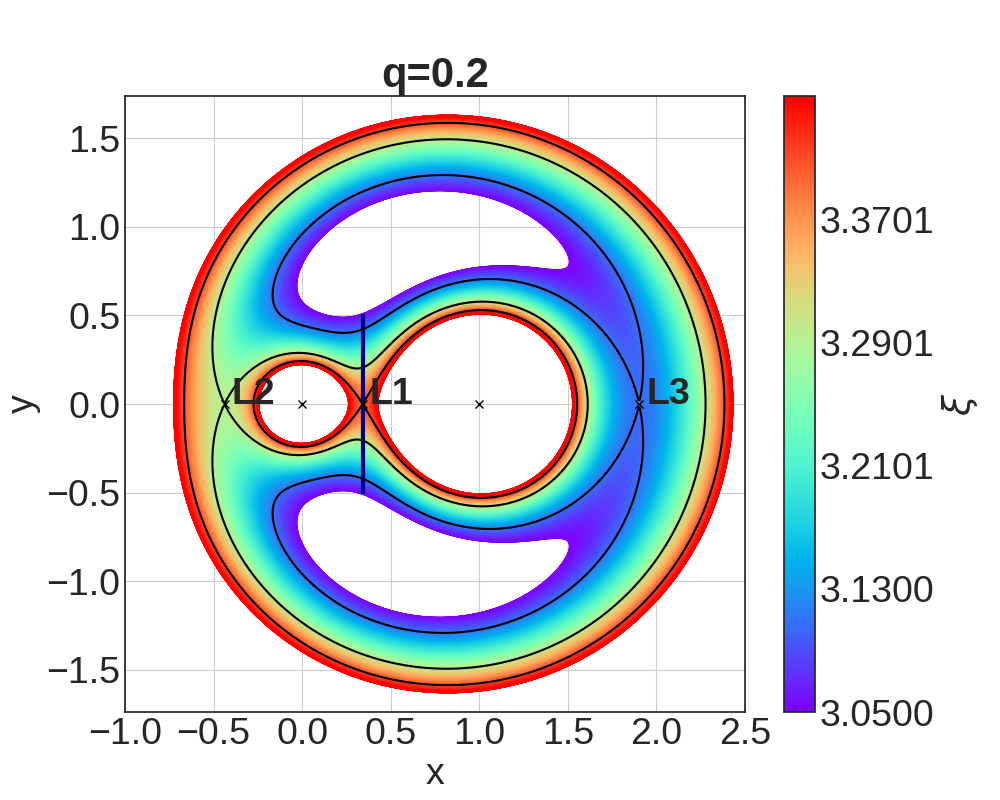}
\includegraphics[width=\columnwidth]{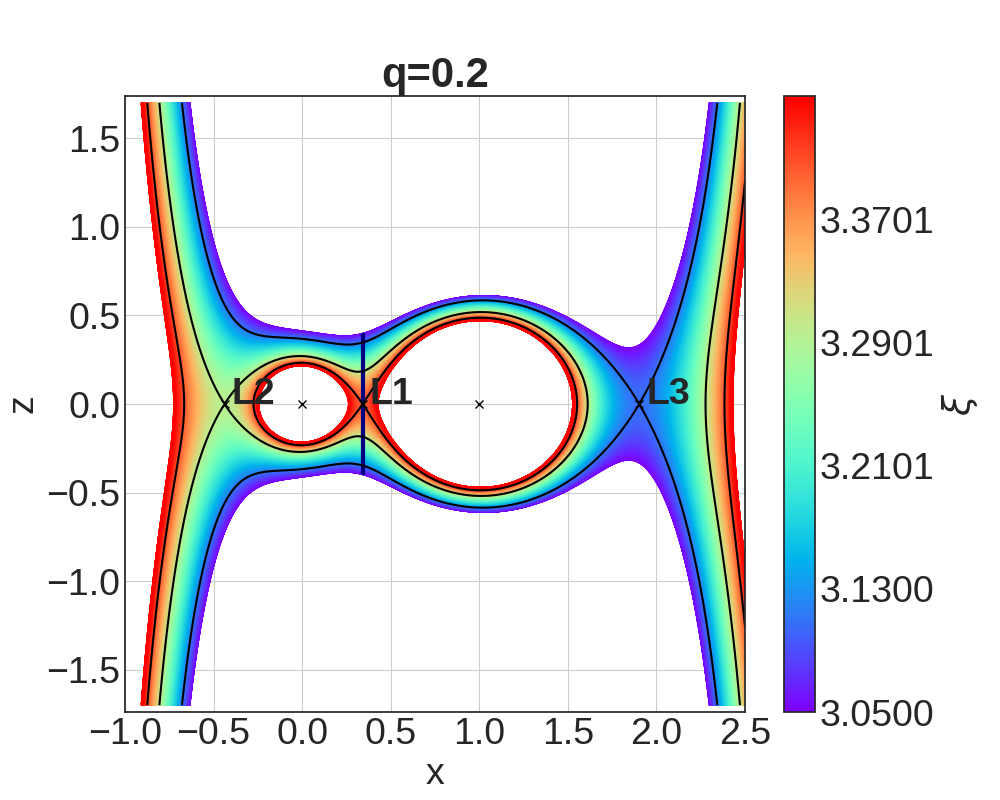}
\includegraphics[width=\columnwidth]{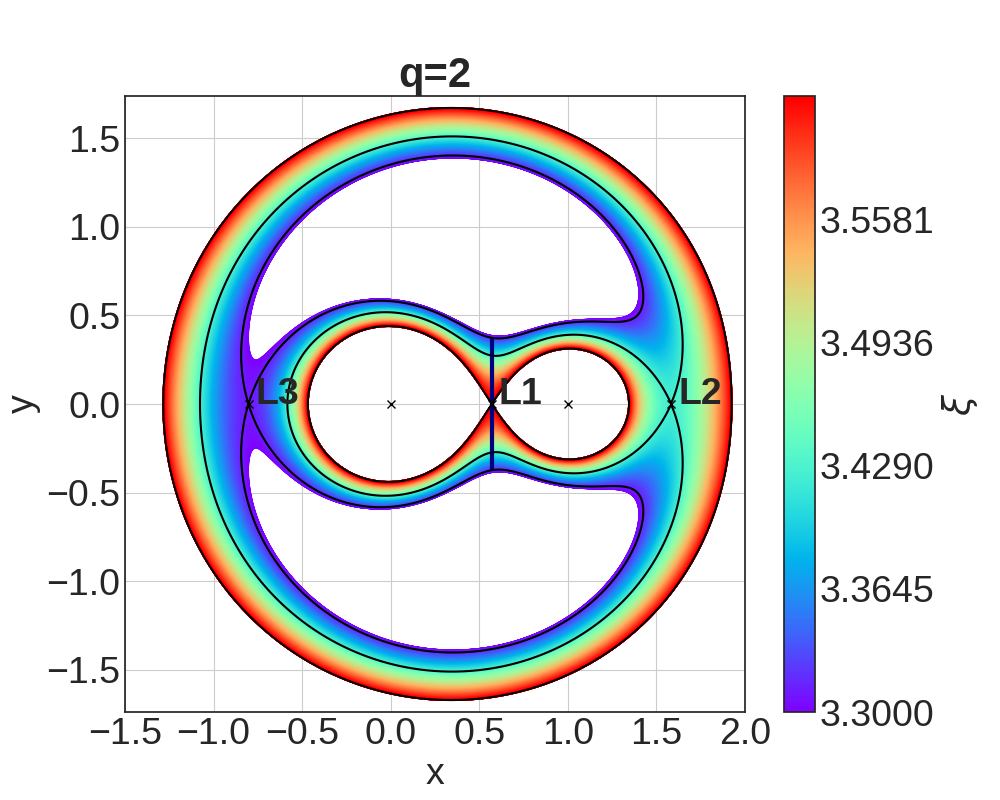}
\includegraphics[width=\columnwidth]{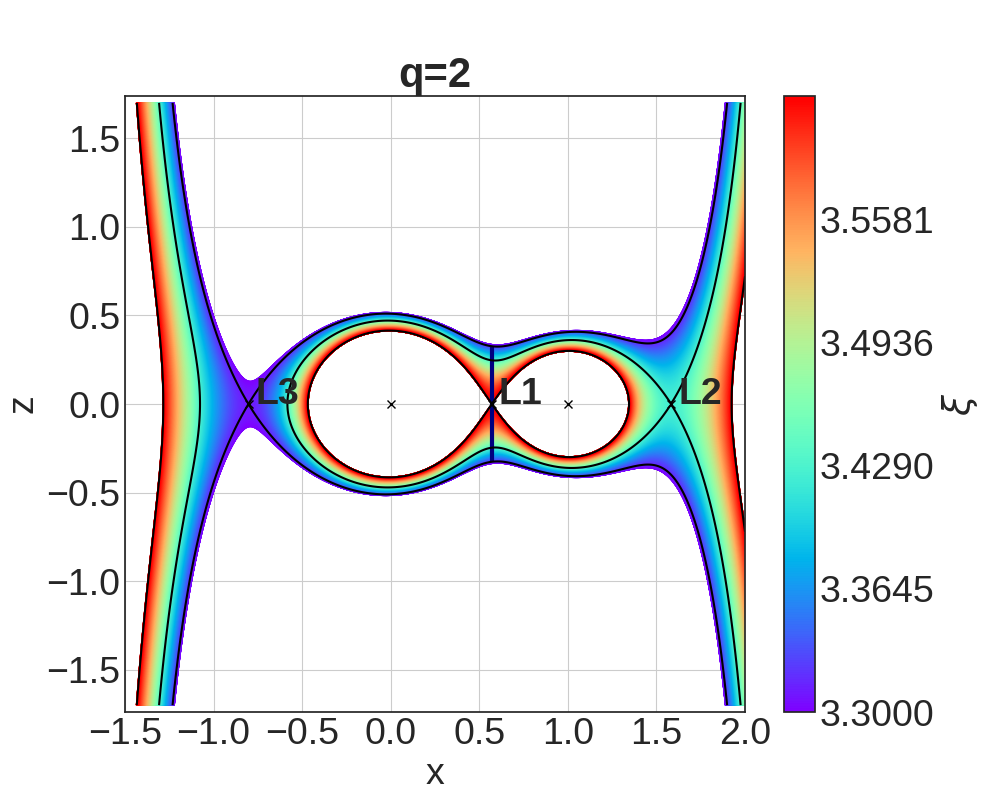}
\includegraphics[width=\columnwidth]{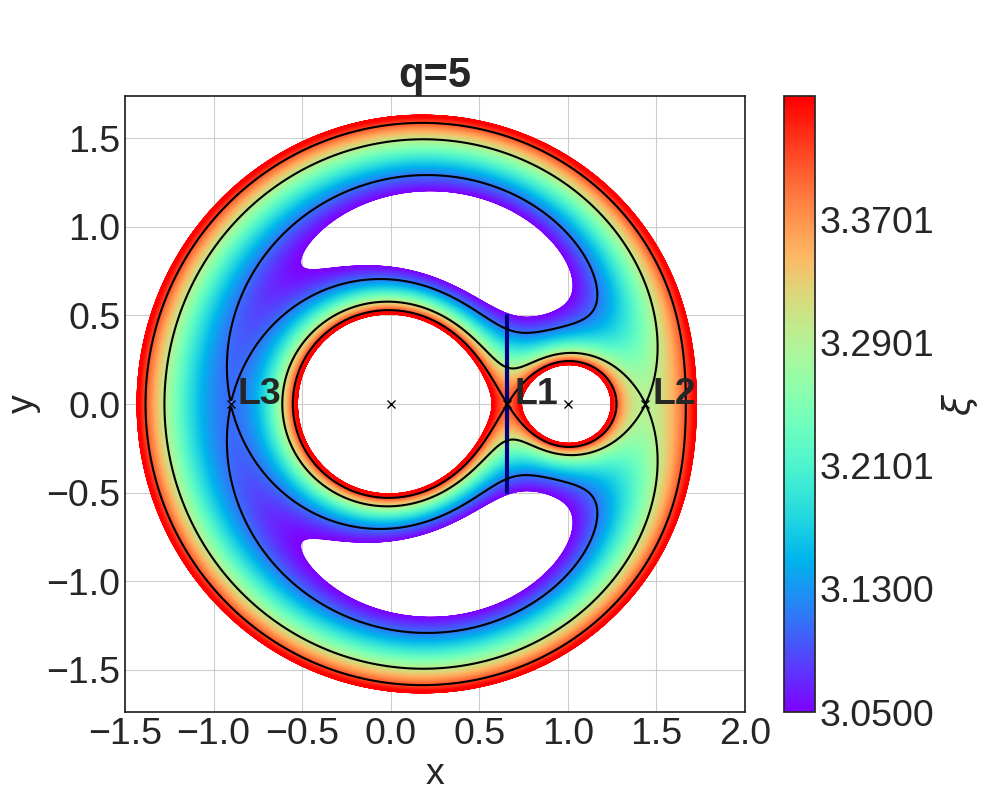}
\includegraphics[width=\columnwidth]{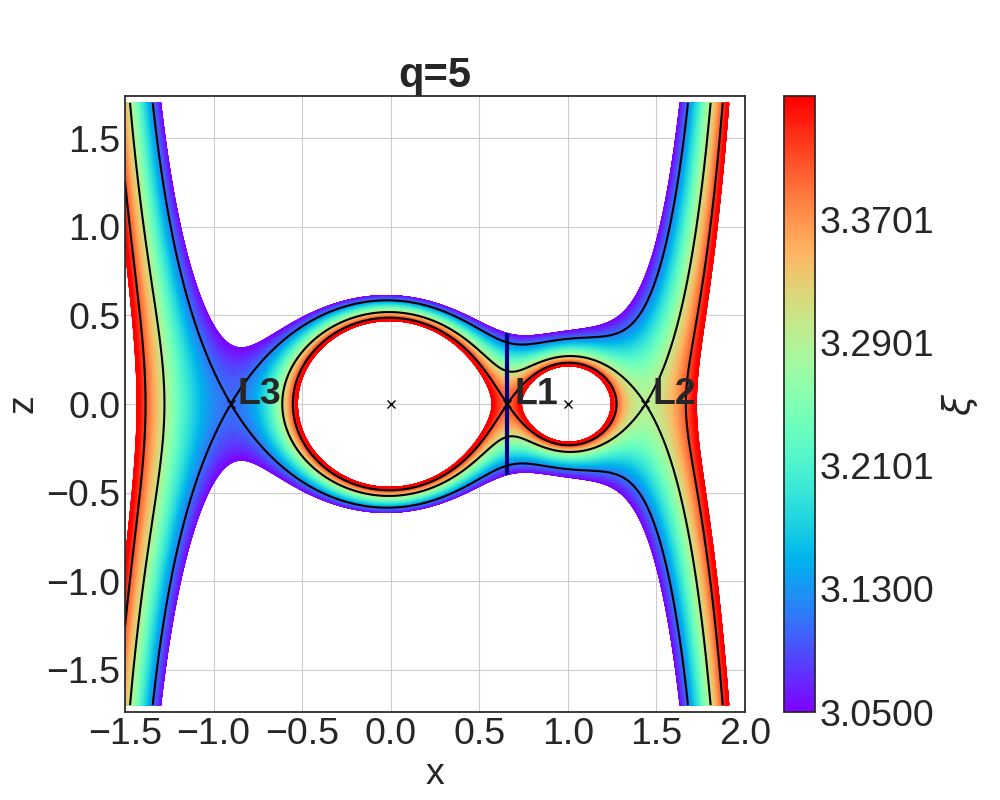}
\caption{Two-dimensional slices showing the scaled potential $\xi$ in a binary corotating frame for three mass ratios: $q=0.2, 2$, and $5$, from top to bottom. Shown are slices in the $xy-$plane (left panels) and $xz-$plane (right panels).  The color bar represents the chosen range of the scaled potential $\xi$ at each point. We plot a narrow range of equipotentials for a better resolution in the region of interest; the white areas are where $\xi$ is outside the chosen ranges. The coordinates are in units of the orbital separation, and the Lagrange points have been denoted by $Li$. The black line passing through  $L_1$ is the $L_1$-plane (the plane through the first Lagrange point and perpendicular to the line attaching the two stars). The equipotential surfaces passing through $L_1$,  $L_2$, and $L_3$ are shown with black lines.
}
\label{fig:2drochelob}
\end{figure*}

We consider a coordinate system in corotation with the binary, with the origin at the donor star's center. The $xy$ plane is the plane of the orbit, and the co-rotating plane is rotating around the axis which passes through the center of mass of the binary system. 
The effective potential $\Psi$ is then

\begin{eqnarray}
    \Psi (X,Y,Z) &=& -\frac{GM_1}{|R_1|}-\frac{GM_2}{|R_2|} \nonumber \\
     & & -\frac{1}{2}\Omega^2\left [(X-a\frac{M_2}{M_1+M_2})^2+Y^2 \right ] \ ,
\end{eqnarray}
\noindent where $M_1$ and $M_2$ are the donor and companion star masses, $a$ is the orbital separation, $\Omega=\sqrt{G(M_1+M_2)/a^3}$ is the binary system's orbital angular velocity, and $R_1$ and $R_2$ are the distances of each element to the donor star's and companion star's centers,

\begin{eqnarray}
|R_1|&=&\sqrt{X^2+Y^2+Z^2} \ , \nonumber \\
|R_2|&=&\sqrt{(X-a)^2+Y^2+Z^2}  \ .
\end{eqnarray}

We construct a unitless (or scaled) potential $\xi$ following the convention of  \cite{mochnacki1984accurate},

\begin{equation}
    \Psi\equiv - \frac{G(M_1+M_2)}{2a} \ \xi \ .
\end{equation}

    
We introduce the mass ratio $q$  
\begin{equation}
    q\equiv\frac{M_1}{M_2} \ ,
\end{equation}
    
\noindent and unitless distances    
    
\begin{equation} x\equiv\frac{X}{a},\ y\equiv\frac{Y}{a}, \ z\equiv\frac{Z}{a},  \ |r_1|\equiv\frac{|R_1|}{a} ,  \ |r_2|\equiv\frac{|R_2|}{a} \ .
\end{equation}
    
Then $\xi$ can be written as

\begin{eqnarray}
 \xi (x,y,z) &=& \frac{q}{(1+q)} \ \frac{2}{|r_1|} + \frac{1}{1+q} \  \frac{2}{|r_2|} \nonumber \\  & &+\left [ (x-\frac{1}{1+q})^2+y^2 \right ].   
\end{eqnarray}

Example contour plots of the unitless equipotentials for several mass ratios (two-dimensional slices for $xy$ plane and $xz$ plane) are shown in Figure~\ref{fig:2drochelob}. In the case of a binary considered in a corotating frame, five equilibrium points exist that correspond to local extrema of the effective potential. The most important for studies of binary interactions are the positions of unstable equilibrium $L_1$, $L_2$, and $L_3$, located on the line through the centers of the two large bodies (see examples shown in Figure~\ref{fig:2drochelob}). In this paper, we will refer to $L_1$ as the inner Lagrangian point and $L_2$ and $L_3$ as the outer Lagrangian points. The outer Lagrangian point closest to the donor star is $L_2$ for $q<1$, and is $L_3$ for $q>1$.

\subsection{Effective acceleration.}
  
\begin{figure}
\includegraphics[width=\columnwidth]{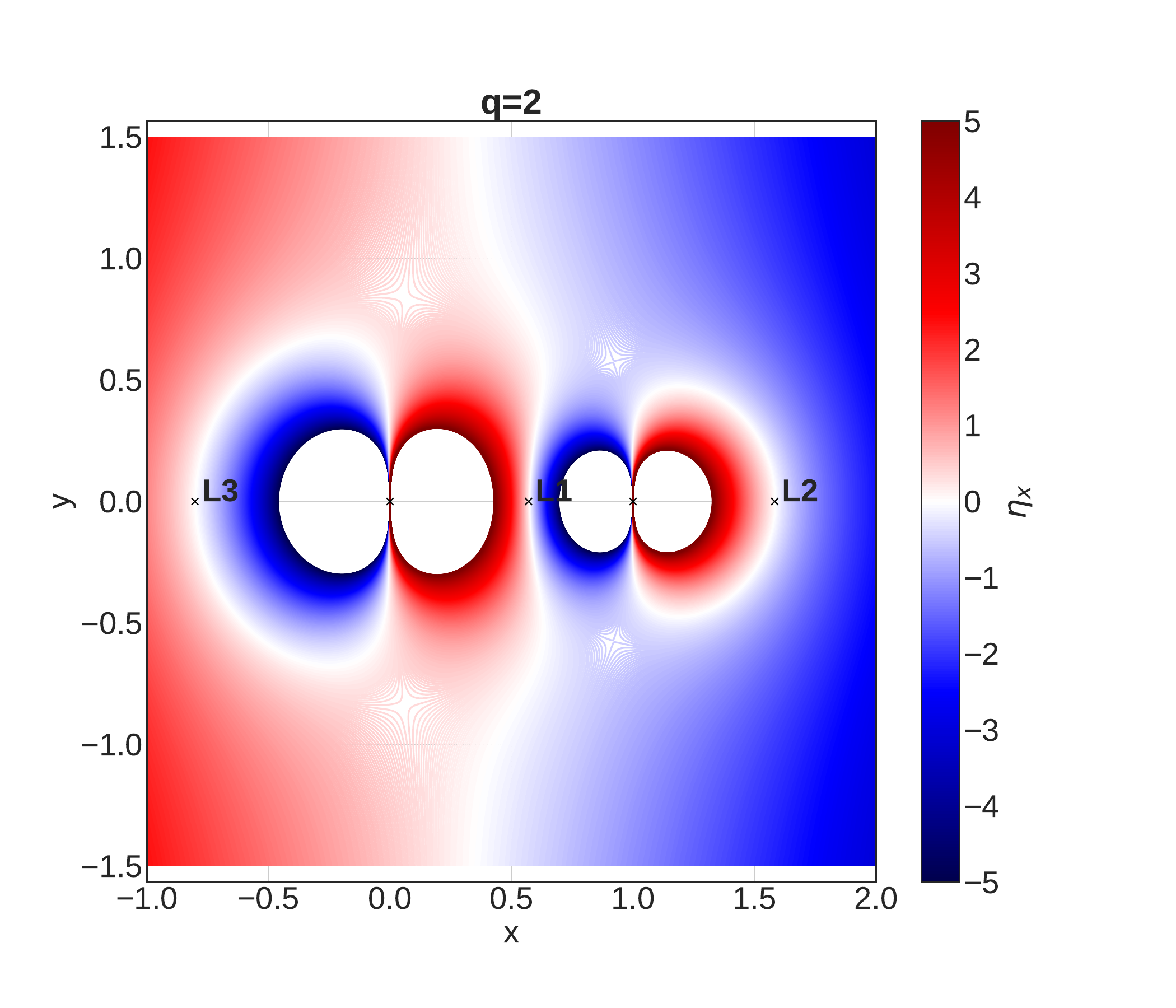} 
\includegraphics[width=\columnwidth]{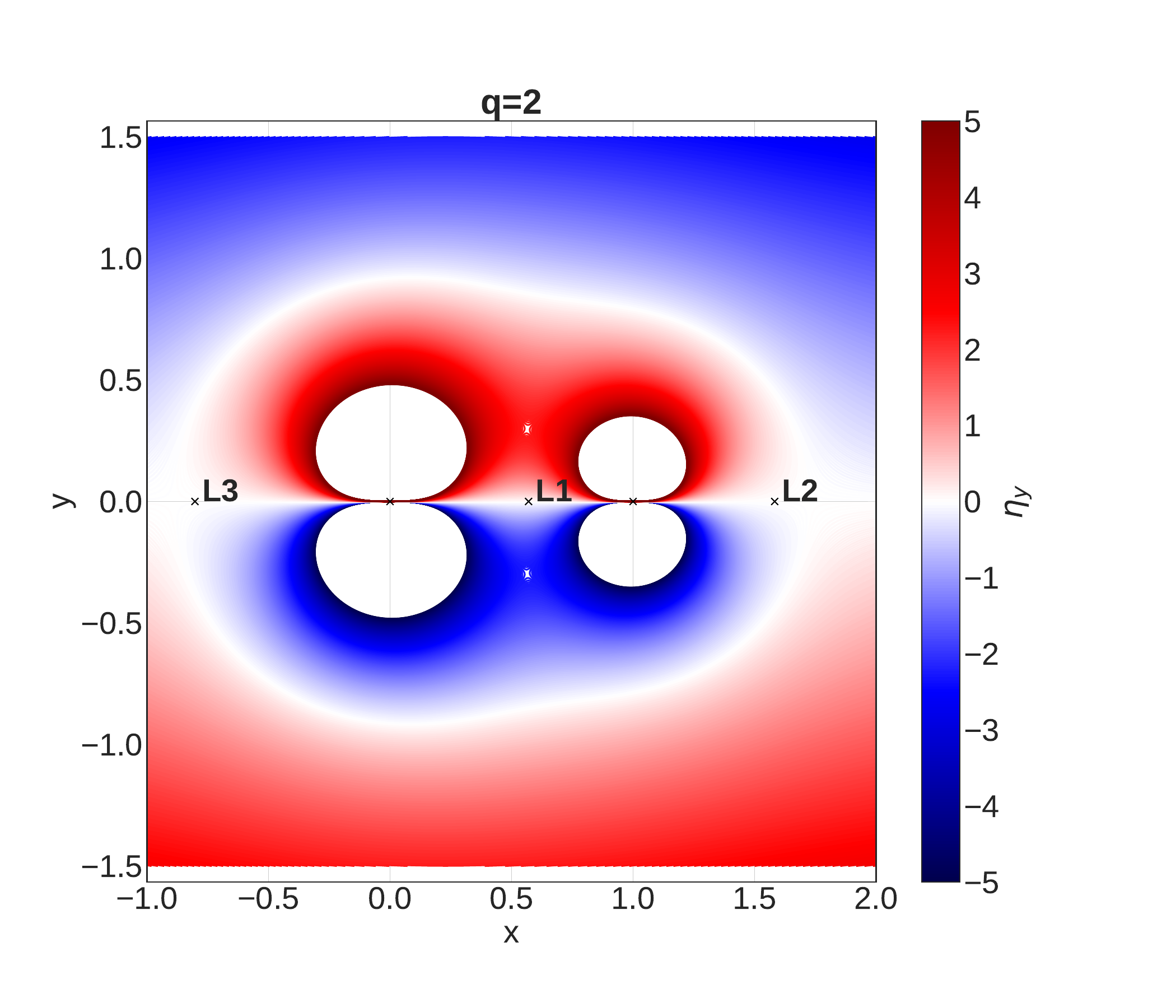} 
\includegraphics[width=\columnwidth]{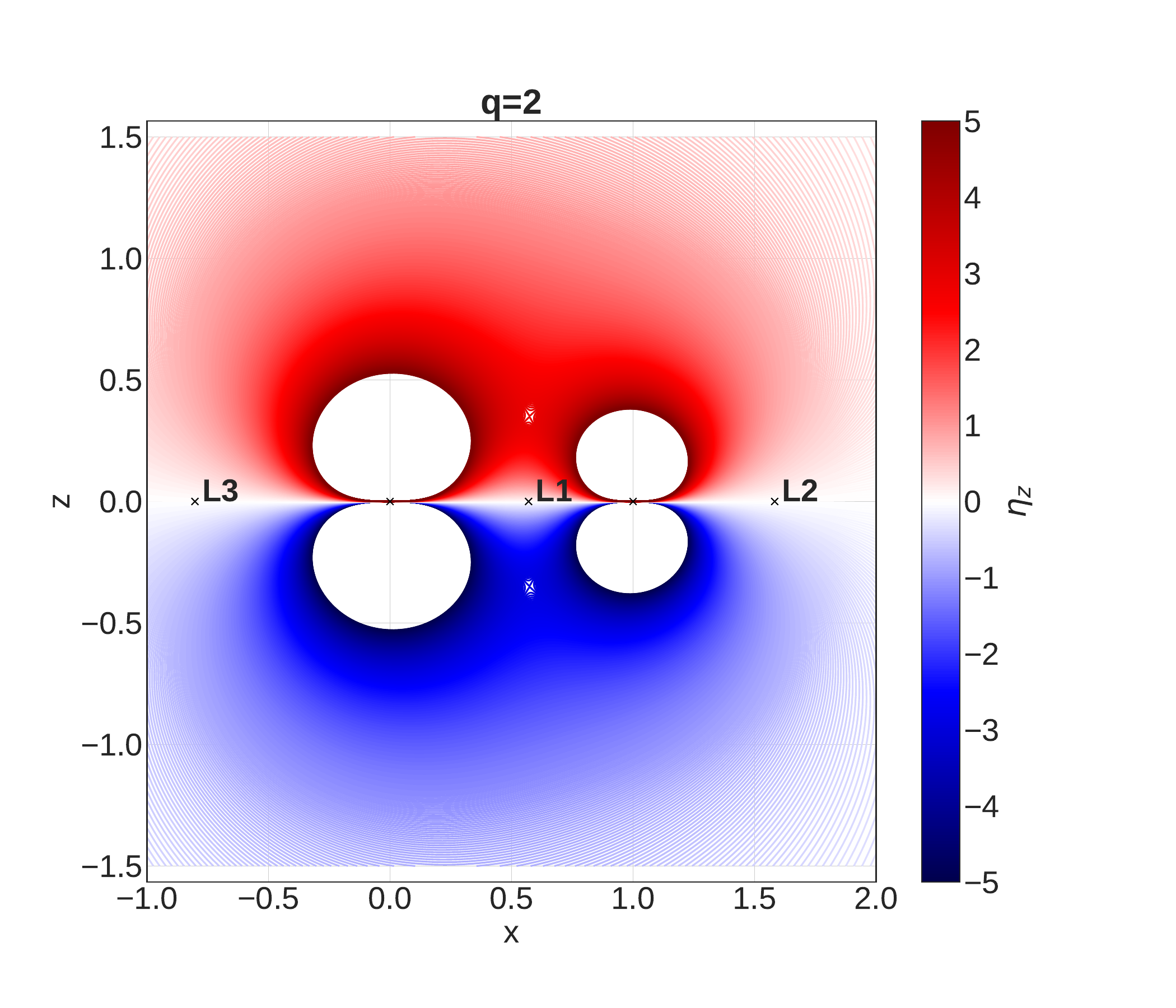} 
\caption{$\eta_x$, $\eta_y$, and $\eta_z$ shown for mass ratio $q=2$. The coordinates are in units of the orbital separation. }
\label{fig:eta_dir}
\end{figure}  
  
We are interested in finding the unitless effective acceleration $\eta$, acting in the direction normal to the equipotential surfaces. This is the gradient of the potential, which is perpendicular to the equipotential surface at each point,

\begin{equation}
    \label{eq:grad}
    \eta=-\nabla \xi.
\end{equation}

 It is easiest to find this gradient of the potential by using its $x,y,$ and $z$ components,  $\eta_x$, $\eta_y$, and $\eta_z$, respectively,\\
\begin{eqnarray}
\label{eq:etax}
    \eta_x&=&  \frac{2q}{1+q} 
    \frac{x  \ |r_1|}{r_1^4}
    + \frac{2}{1+q} \frac{ (x-1)\ |r_2| }{r_2^4} \nonumber \\
    & & 
    - 2\left [x-\frac{1}{1+q} \right] \ , \\ 
\label{eq:etay}
    \eta_y&=& \frac{2q}{1+q}\frac{y \ |r_1|}{r_1^4} + 
    \frac{2}{1+q}\frac{y \ | r_2|}{r_2^4} - 2y \ , \\     
\label{eq:etaz}
    \eta_z&=&  \frac{2q}{1+q} \frac{ z \ |r_1|}{r_1^4} +
    \frac{2}{1+q}\frac{z \ |r_2|}{r_2^4} \ .
\end{eqnarray}

\noindent A plot depicting  $\eta_x$, $\eta_y$, and $\eta_z$ can be seen in Figure~\ref{fig:eta_dir}. 
The unitless effective acceleration $\eta$ is then given by

\begin{equation}
\eta=\sqrt{\eta_x^2+\eta_y^2+\eta_z^2} \ .
\label{eq:tot_eta}
\end{equation}

\begin{figure*}[p]
  \centering
\includegraphics[width=\columnwidth]{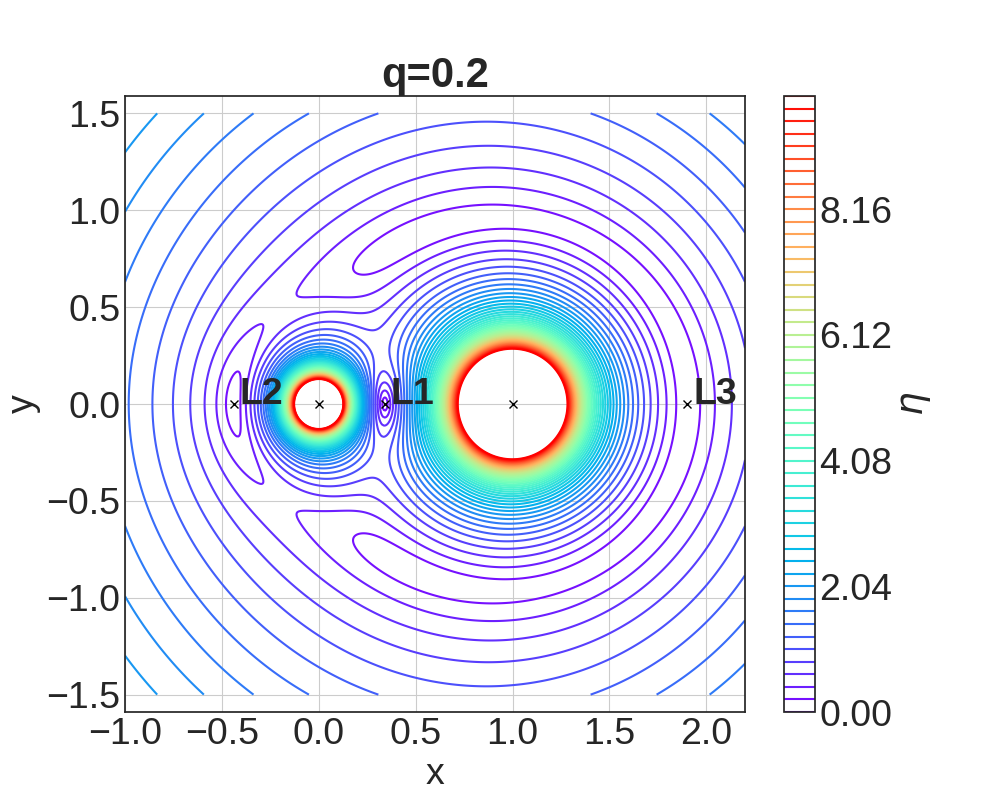}
\includegraphics[width=\columnwidth]{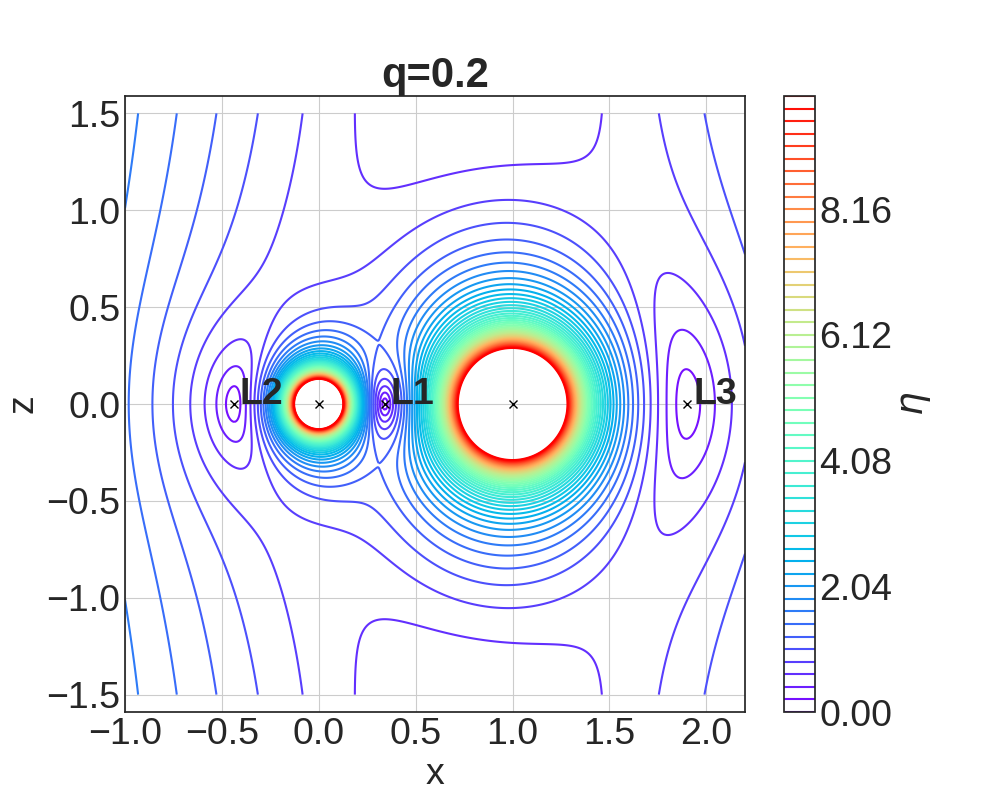}
\includegraphics[width=\columnwidth]{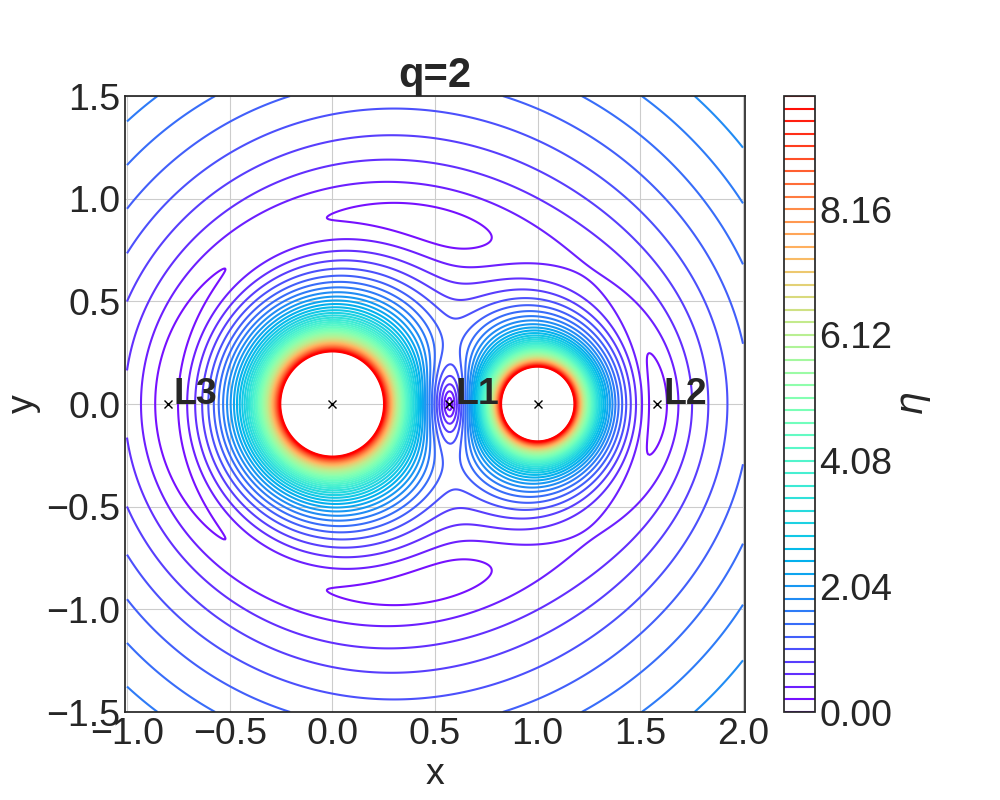}
\includegraphics[width=\columnwidth]{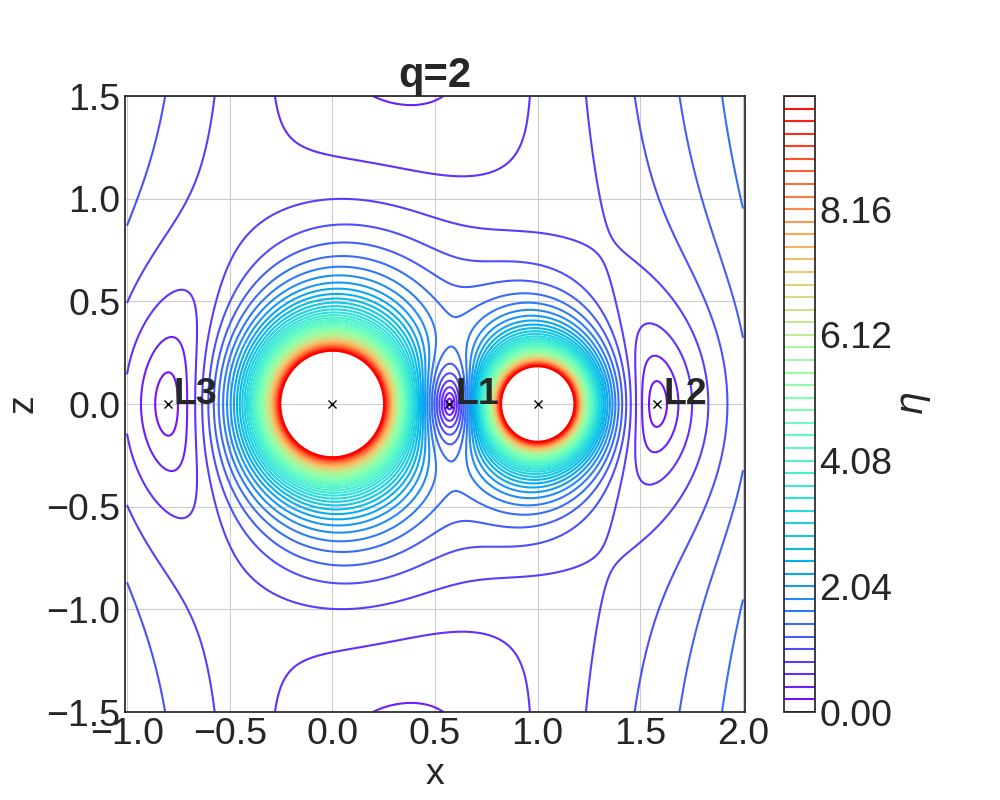}
\includegraphics[width=\columnwidth]{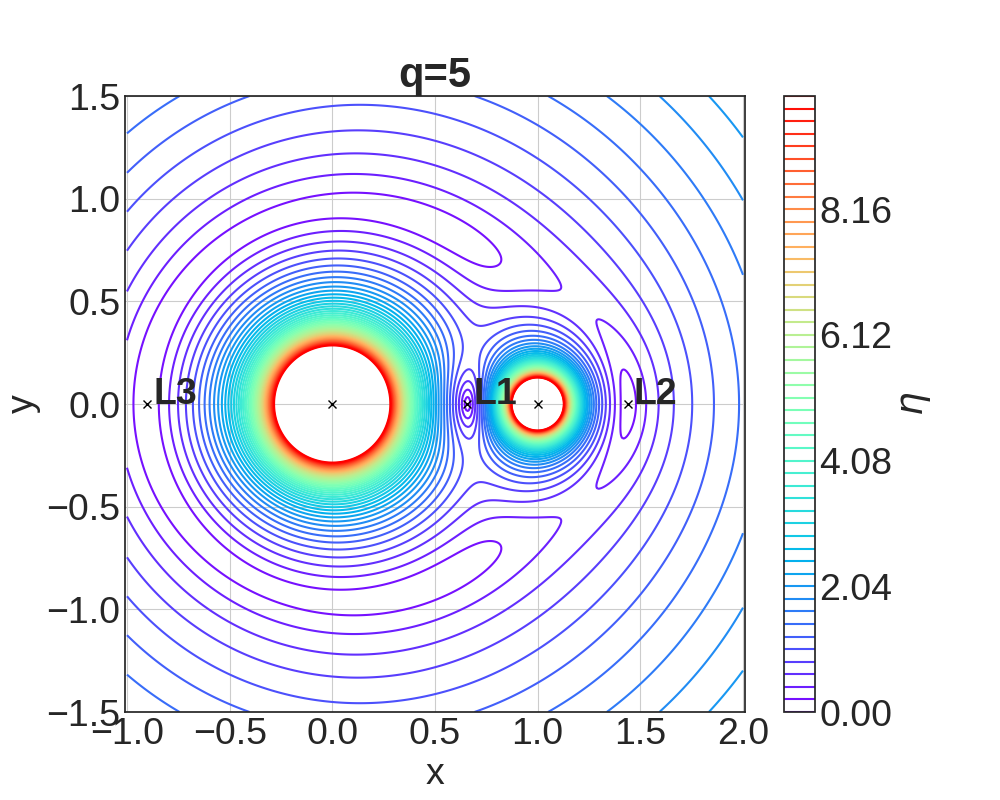}
\includegraphics[width=\columnwidth]{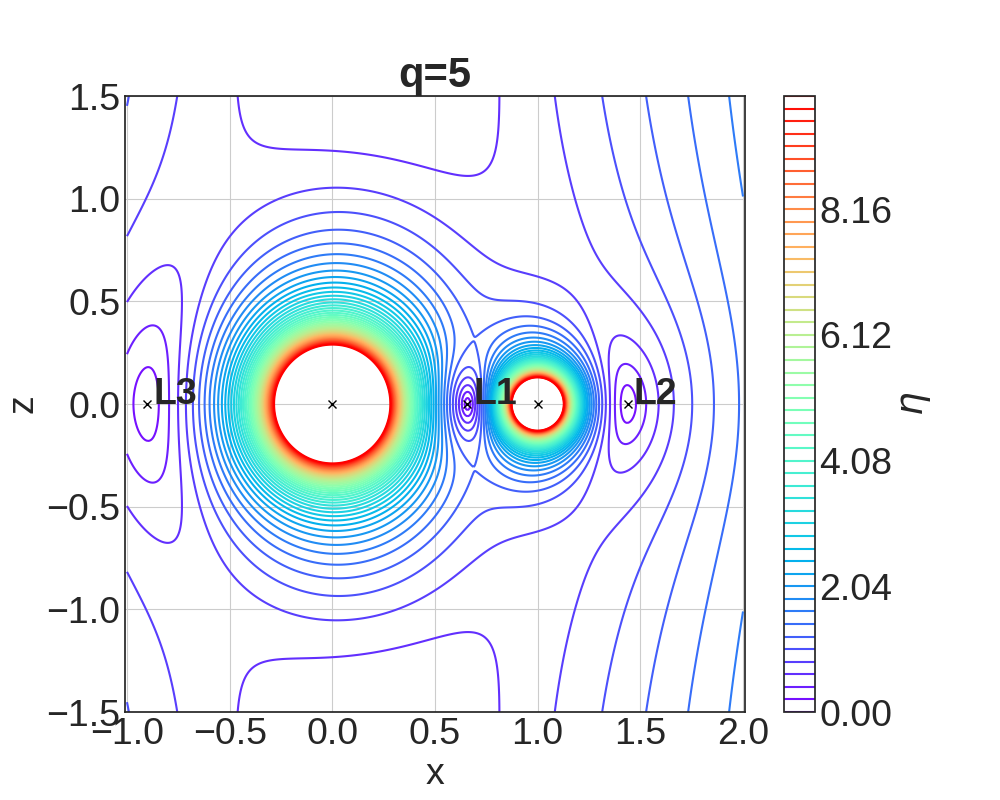}
\caption{Two-dimensional slices showing the contours of the effective acceleration $\eta$, in a binary corotating frame, for three mass ratios: $q=0.2$, $2$, and $5$, from top to bottom.  Shown are slices in the $xy-$plane (left panels) and $xz-$plane(right panels). The coordinates are in units of the binary separation, and the Lagrange points have been denoted by $Li$. }
\label{fig:2dacc}
\end{figure*}  

\noindent Contours for effective accelerations are shown  in Figure~\ref{fig:2dacc}.
The effective acceleration with physical units $g_{\rm eff}=$ can be then recovered as:
\begin{equation}
    g_{\rm eff}= \frac{G(M_1+M_2)}{2a^2}\eta.
\label{eq:scalegrav}
\end{equation}

\section{One-dimensional representation}

As shown in Figures \ref{fig:2drochelob} and \ref{fig:2dacc}, for stars in a binary system, there is no spherical symmetry for neither effective potentials nor effective accelerations. Close to the center of each star, a close-to-spherical symmetry can be observed, but the further away from the center of each star one goes, the more significant the influence of its companion and of the orbital motion becomes. The closer one gets to the $L_1$ point, the more the shape of the potential gets close to a teardrop shape. Outside of the $L_1$ point, the equipotentials take on a shape similar to a peanut. The effective acceleration, on the other hand, entirely disappears at $L_1$.

\begin{figure}[ht!]
	\centering               
	\includegraphics[width=\columnwidth]{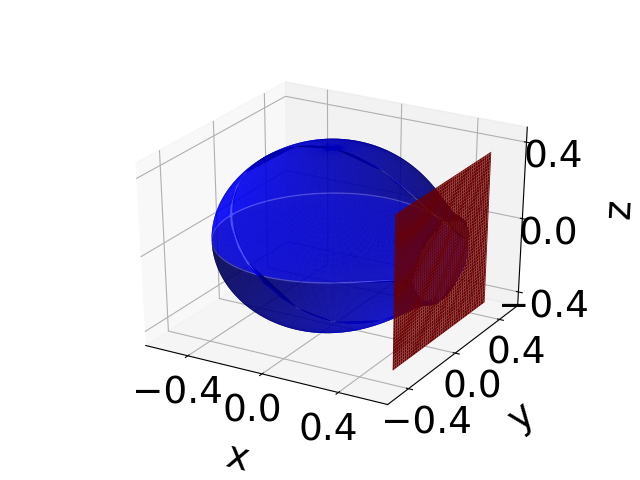}
	\caption{A sample equipotential surface outside of the Roche lobe in a binary star, shown in blue color. The equipotential surface is limited by the $L_1$-plane, shown in red color. The coordinates are in  units of the orbital separation.}
		\label{fig:3D shell}
\end{figure}

Our goal here is to develop a method to approximate 3D binary stars' gravitational fields for 1D considerations. Our approach is based on the concept of equipotential shells (3D shells that enclose the center of the stars). We will only focus on the donor star for the remainder of the paper. A 3D depiction of a sample equipotential shell outside of the Roche lobe and limited by the $L_1$-plane can be seen in Figure~\ref{fig:3D shell}. The $L_1$-plane is the plane that is parallel to the $y-z$ plane, and passes through the $L_1$ point.

The key physical property of the equipotential shells is that, in the case of hydrostatic equilibrium, both isobaric surfaces and surfaces of constant density coincide with equipotential surfaces. 
Therefore here we consider averaging the effective acceleration that affects the star over the equipotential surfaces. The effective accelerations are found as a function of the volume-equivalent radii of the same equipotential surfaces.

\subsection{Volume-equivalent radii}

\label{sec:Volume-equivalent-radii}

Each equipotential surface encloses some volume $V_{\rm equip}$. This volume can be associated with a sphere of the same volume. This sphere has a so-called volume-equivalent radius

\begin{equation}
    R_{\rm eq}=\sqrt[3]{\frac{3V_{\rm equip}}{4\pi}} \ .
        \label{eq:vol_equiv_radius}
\end{equation}

The idea of the volume-equivalent radius was introduced in the past \citep[see, for example, ][]{kopal1959close}. The best known numerical fitting of 3D integrations for the unitless volume-equivalent Roche lobe radius is the famous equation of  \citet{eggleton1983approximations}

\begin{equation}
    r_{L_1}=\frac{0.49q^{2/3}}{0.6q^{2/3}+\ln(1+q^{1/3})}\ .
        \label{eq:eggleton}
\end{equation}

\noindent The physical volume-equivalent Roche lobe  radius of the star $R_{L_1}$ then is found by multiplying $r_{L_1}$ by the orbital separation $a$, $R_{L_1} = r_{L_1} a$.

For volume integrations, we use spherical coordinates.  We define $i$ and $j$ as the indices describing the directions in $\varphi$ (azimuthal angle) and $\vartheta$ (polar angle) for integrations, where $\varphi\in[0,\pi]$ and $\vartheta\in[0,{\pi}/{2}]$\footnote{The values of unitless acceleration, $\eta$, and equipotentials, $\xi$, in a binary system have the symmetry along $z$-direction with respect to the orbital plane $xy$, as well as along $y$-direction with respect to  $xz$-plane. Therefore, only integrations in $\varphi\in[0,\pi]$ and $\vartheta\in[0,{\pi}/{2}]$ are needed.}. We divide each of these domains into $n_{\phi}$ and $n_{\vartheta}$ respectively (we adopt $n_\varphi=2 n_{\vartheta}$). We define $k$ to be the index describing the potential shell with $\xi_k$. The $r_{ijk}$ is the location of the equipotential shell $\xi_{k}$ in each angular direction $ij$, or, specifically, the distance to the origin $r$ for the given angles $\varphi$ and $\vartheta$. $r_{ijk}$  is found iteratively until at least the convergence condition 

\begin{equation}
    \left |\frac{\xi(r_{ijk},\varphi_i,\vartheta_j)-\xi_{k}}{\xi_{k}}\right |\leq 10^{-12} 
    \label{eq:prec_condition}
\end{equation}

\noindent is satisfied. Here $\xi(r,\varphi,\vartheta)$ is the potential at the location $(r,\varphi,\vartheta)$, while $\xi_{k}$ is the potential for the equipotential surface for which we search.

The numerical integrator uses spherical volume elements in each angular direction $\varphi$  and $\vartheta$, and the volume between the $k_{th}$ shell and $(k-1)_{th}$ shell in the potential is

\begin{equation}
    \Delta V_{\rm{sph},k}= r_{ijk}^2 \sin{\vartheta_j} \Delta r_{ijk} \Delta \varphi \Delta \vartheta \ .
    \label{eq:dvsph}
\end{equation}

\noindent $\Delta r_{ijk}$   is the step in the $ij$ direction. This is not constant for the entire potential shell $k$ but depends on the location of the currently sought potential $\xi_k$ in $ij$-direction, and the location of the previous potential, $\xi_{k-1}$. 
Using the symmetry of the equipotential in a binary system, we perform the volume integration in the angular domain as above and then multiply by 4.

The volume integrations are limited by the outer Lagrange point's equipotential shell of the donor star, and the $L_1$ plane. These limitations affect volume integrations for shells that are located outside the $L_1$ equipotential. The $L_1$ plane is depicted as a black line in the 2D slice of Figure~\ref{fig:2drochelob} and as a red plane in the 3D Figure~\ref{fig:3D shell}.

\subsection{Effective accelerations}

\begin{figure}[ht!]
	\centering        
	\includegraphics[width=\columnwidth]{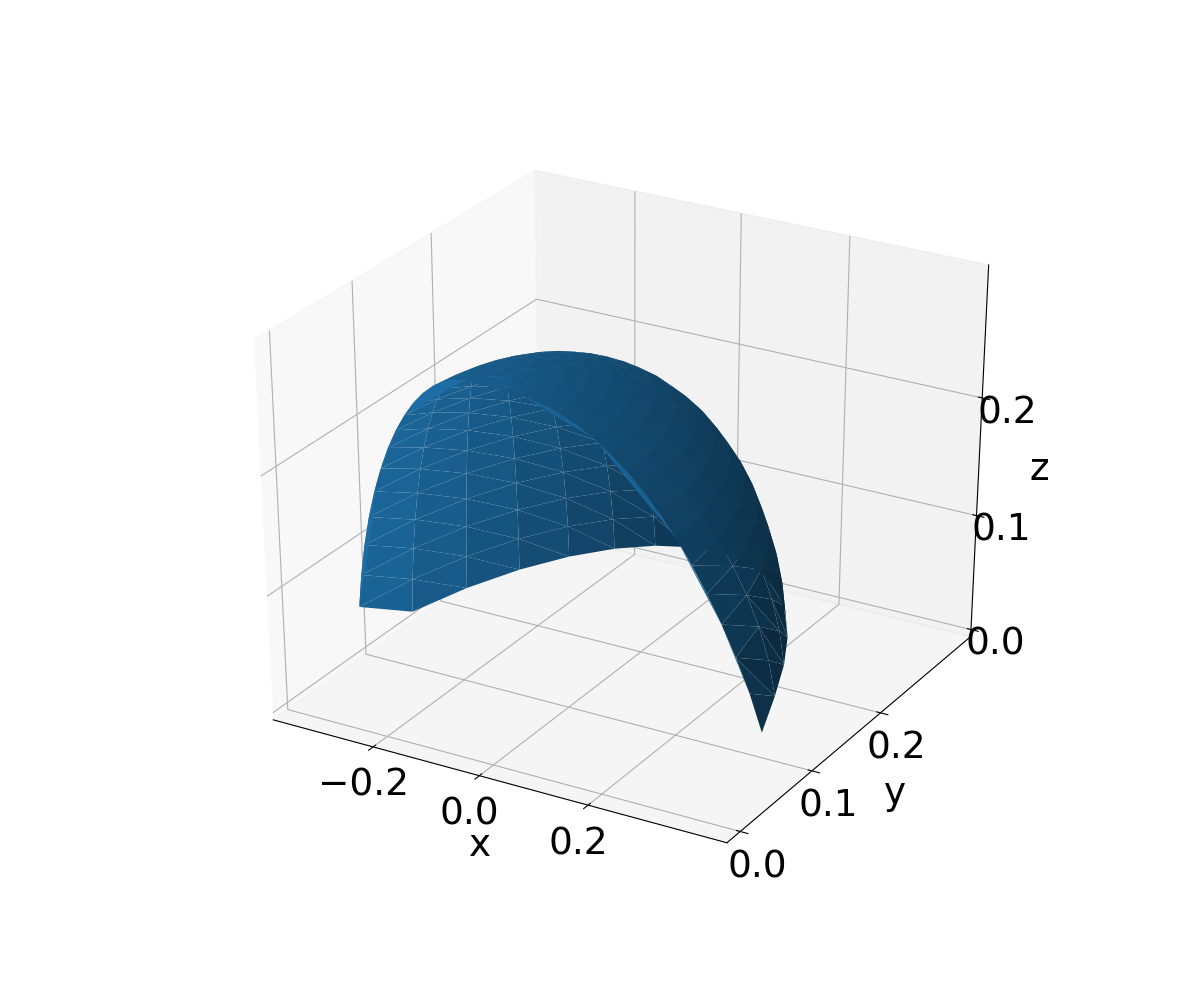}
	\caption{A depiction of a part of the  equipotential shell with triangular meshes on it shown in 3D; for a random shell in the case of the mass ratio $q=0.5$. Only a part of the integration domain is shown, to make the triangular mesh visible. }
\label{fig:sample_mesh}
\end{figure}

The unitless effective accelerations at each location are found using equations \ref{eq:etax}, \ref{eq:etay}, and \ref{eq:etaz}. 

The weighted averages of the effective acceleration for each equipotential shell are obtained with the area of the triangular elements on that equipotential surface $dA_{ij}$\footnote{We compared the performance of triangular mesh area elements and spherical area elements \citep[as did ][]{mochnacki1984accurate} and found that the triangle method provides faster convergence to obtain surfaces on equipotential shells. This is especially important for the shells outside  the Roche lobe, where element areas of the surfaces of equipotentials do not resemble spherical area elements.},

\begin{equation}
    \eta_{\rm avg} = 
    \frac{ \sum_{i} \sum_{j}  \eta_{ij} dA_{ij}  }
    {\sum_{i} \sum_{j} dA_{ij} } \ ,
\label{eq:weighed_avg_acc}
\end{equation}

\noindent where $i$ and $j$ are the indices of $\varphi$ and $\vartheta$ for which the triangles are summed. Here
$\eta_{ij}$ is the unitless effective acceleration averaged over three triangles' vertices.

Figure~\ref{fig:sample_mesh} shows a sample triangular mesh scheme for an equipotential shell. 

\subsection{Integration zones}

 \begin{figure}[t!]
	\centering                    
	\includegraphics[width=\columnwidth]{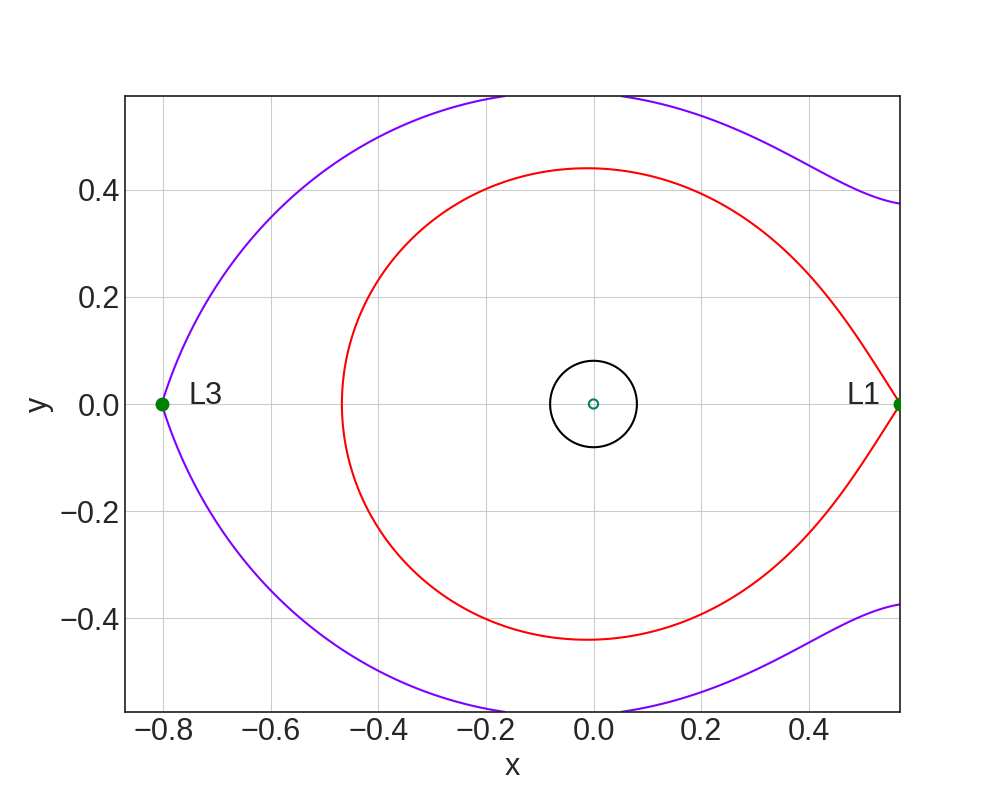}
	\caption{ The three zones of integration; this specific example is for a mass ratio of $q=2$. The zones are defined as described in equation~\ref{eq:integr_ranges}, see \S~3.3 for further details.}
		\label{fig:ranges}
\end{figure}

The potential is very sensitive to the distance from the center. To accommodate this sensitivity, we divide the integration zones into three regions: 

\begin{eqnarray}
\xi_1&=&\xi(0,0,0,q)  \\
\xi_2&=&\xi(0.05 x_{L_1},0,0,q) \nonumber \\
\xi_3&=&\xi(x_{L_1},0,0,q)\nonumber \\
\xi_4&=&\xi(x_{L,\rm outer},0,0,q) \nonumber 
\label{eq:integr_ranges}
\end{eqnarray}
 \noindent An example of the three integration regions can be seen in Figure~\ref{fig:ranges}.  
 
Near the center, the true binary potential and effective acceleration $\eta$ are not very different from these in a spherically symmetric single star

\begin{eqnarray}
\xi &= &\xi_{\rm ss} + \Delta \xi , \quad  
{\Delta \xi} = \left ( \frac{3+2q}{2q(1+q)} r + \frac{2q-1}{6q} r^3 \right )  \xi_{\rm ss} \ , \nonumber \\
\eta &= &\eta_{\rm ss} + \Delta \eta , \quad 
{\Delta \eta} = - \frac{2}{3} \frac{ (1+q)}{q}  r^3 \eta_{\rm ss}\ .
\label{eq:center_summary}
\end{eqnarray}

\noindent Here $\xi_{\rm ss}$ is the unitless effective binary potential in a a spherically symmetric star, $\eta_{\rm ss}$ is the unitless effective acceleration in a spherically symmetric star, and $r$ is the unitless distance to the center of the first star (distance in the units of the binary separation). For the derivation, see derivations in Appendix \S\ref{app:near_center} and Equation~\ref{eq:deviation_close_tocenter}. 
The expected deviations near the center, $\Delta \xi$ and $\Delta \eta$, are  
very small. 

The expected local truncation error due to the limited resolution during numerical integration for any quantity is of the order of $2 \pi^2/n_\phi^2$ times that quantity. We have checked and verified numerically that our integrations for volume equivalent radii and averaged effective accelerations are limited by the adopted angular resolution, as expected for $R_{\rm Err}$. The limited numerical precision with which we can obtain solutions near the center does not allow us to resolve this behavior correctly, for instance the error in finding volume equivalent radii is $\delta_{\rm err}(R_{eq}) \approx (2/3) (\pi^2/n_\phi^2)$ (assuming that $R_{\rm eq}$ is of the order of one). However, this error might be comparable to the radius itself and hence leads to an even larger error in finding the effective gravity. We introduce as $r_{\rm am}$ the unitless distance to the origin within which the analytical solutions using Equation \ref{eq:center_summary}  have to be used, as opposed to the distances $r>r_{\rm am}$, where numerical integrations have to be used.

We have verified that for the mass ratios in the range of $[10^{-6},10^5]$, at $r_{\rm an}=0.05 x_{L_1}$, the non-sphericity effect is $\Delta\xi/ \xi_{\rm ss}\la 10^{-4}$, and hence the higher terms do not contribute substantially. 
This makes  Equation \ref{eq:center_summary} directly applicable. 
The numerical precision to find the potential or the effective acceleration, with $n_{\varphi}=4000$, is approximately $ 10^{-6}$. With this resolution, the numerical ''noise'' due to truncation error contributes at less than a percent of the obtained values for various quantities for our smaller mass ratio $q=10^{-6}$, and as little as $0.01\%$ for all mass ratios $q>1$. The numerical noise decreases with $r$, albeit  the role of terms neglected in Equation \ref{eq:center_summary} grows. Therefore $r_{\rm an}=0.05 x_{L_1}$ is a compromise at which we can still trust the analytic solution but can start to trust the numerical solution for a wide range of mass ratios.

For $r<r_{\rm an}$ we therefore consider that the potential and effective acceleration should be provided by Equation \ref{eq:center_summary}. For the intermediate zone, we start our numerical integrations at $r=r_{\rm an}$.

In our runs, we use  500 shells between $\xi_2$ and $\xi_3$, equally spaced in distance between $0.05x_{L1}$ and $x_{L1}$ and 100 shells between $\xi_3$ and $\xi_4$, equally spaced in the logarithm of the potential between $\xi_{L1}$ and $\xi_{L\rm outer}$.

\subsection{Properties in the $L_1$ neighborhood cross-section}

\label{sec:L_1neighborhood}

\begin{figure}[]
	\centering                      
	\includegraphics[width=\columnwidth]{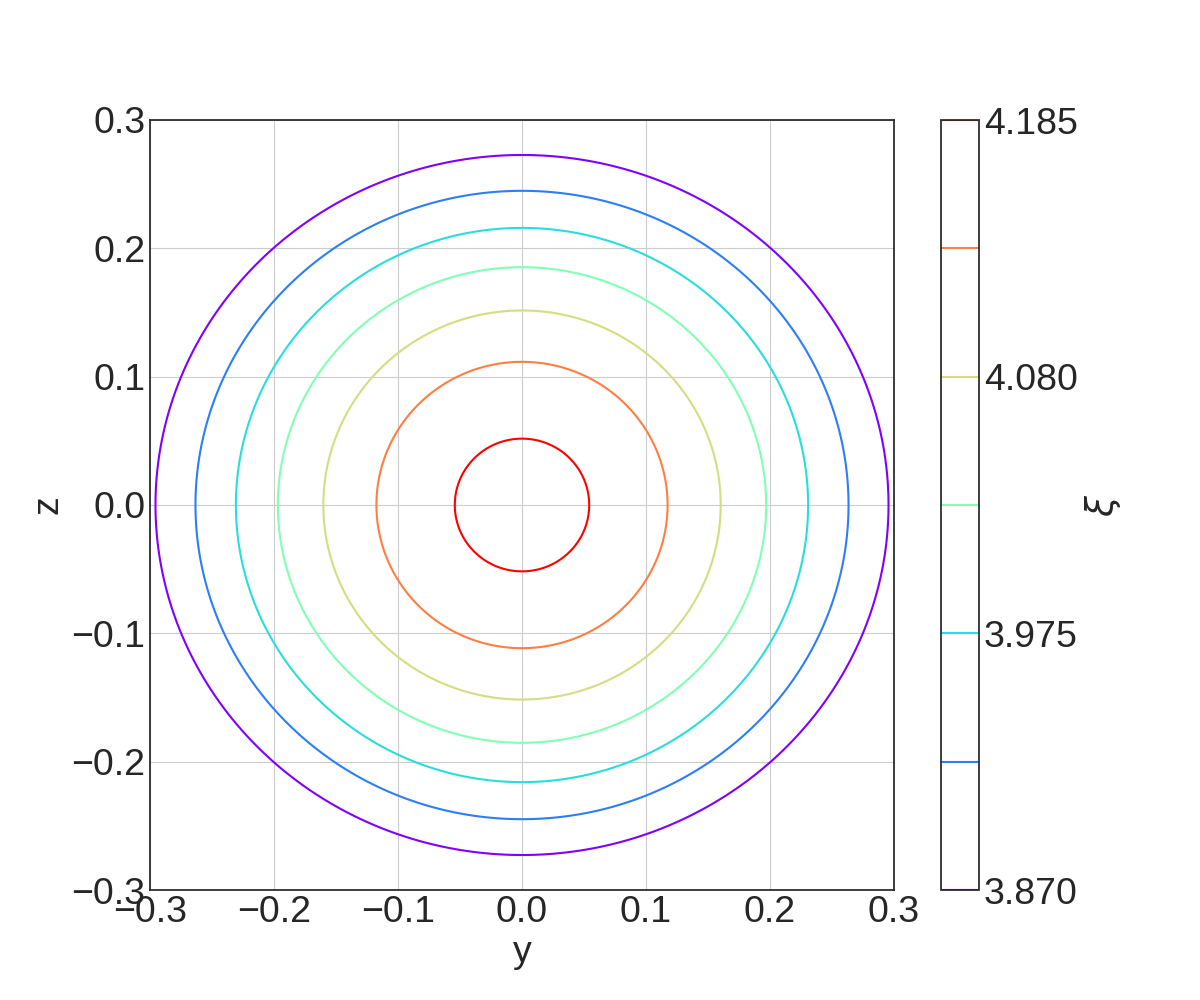}
	\caption{The intersection of several equipotential surfaces with the $L_1$ plane for $q=0.2$. Axes are in units of the orbital separation.}
	\label{fig:L_1plane}
\end{figure}

We also compute several properties at the $L_1$ plane to provide in our database. These properties are functions of unitless equipotential shells passing through the $L_1$ plane. The intersections of equipotential shells with the $L_1$ plane appear as ellipse-like curves on the $L_1$ plane (see Figure~\ref{fig:L_1plane}). The tabulated properties are as follows\footnote{Note that all $L_1$-plane properties are obtained for equipotential shells exceeding the Roche lobe. In the database we provide,  all the columns corresponding to these properties are set to zero for all equipotential shells within the Roche lobe.}:

\begin{itemize}
 \item The area of the cross-section of the $L_1$ plane and the equipotential shell passing it, as a function $\xi$. The area integration uses polar coordinates with 4000 angular resolution per quadrant

  \item The locations of intersections of this cross-section with the $xy$-plane and $xz$-plane, as a function of $\xi$.

  \item The effective acceleration averaged over the intersection between the equipotential and the $L_1-$ plane. 
  
  \item The effective acceleration averaged over the entire area of each of these elliptical cross-sections. 
\end{itemize}

\section{The numerical results}

\label{sec:numres}

\begin{figure}[]
	\centering                     
	\includegraphics[width=\columnwidth]{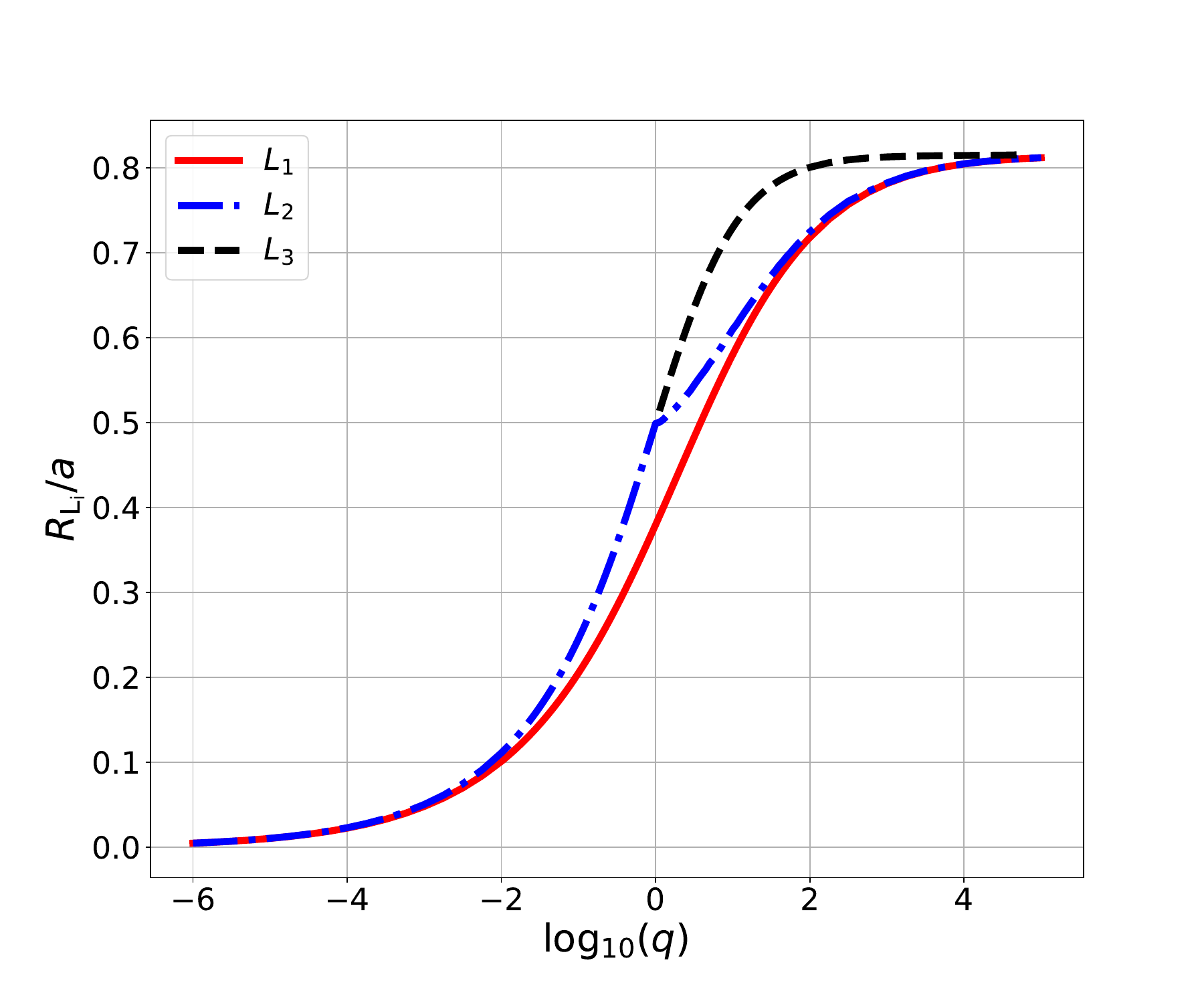}
	\caption{This plot shows the results of our highest resolution run for the volume equivalent radii passing the first three Lagrange points, as a function of the mass ratio. The horizontal axis is in the units of orbital separation. $R_{L3}$ volume around the donor star exists only for mass ratios greater than 1. Note that the outer Lagrange point is $L_3$ for mass ratios greater than 1, and is $L_2$ for mass ratios less than 1.}
	\label{fig:lagrangepoint_voleqrad}
\end{figure}

\subsection{Self-convergence}

\label{sec:converg}

\begin{figure}[ht!]
\centering
\includegraphics[width=\columnwidth]{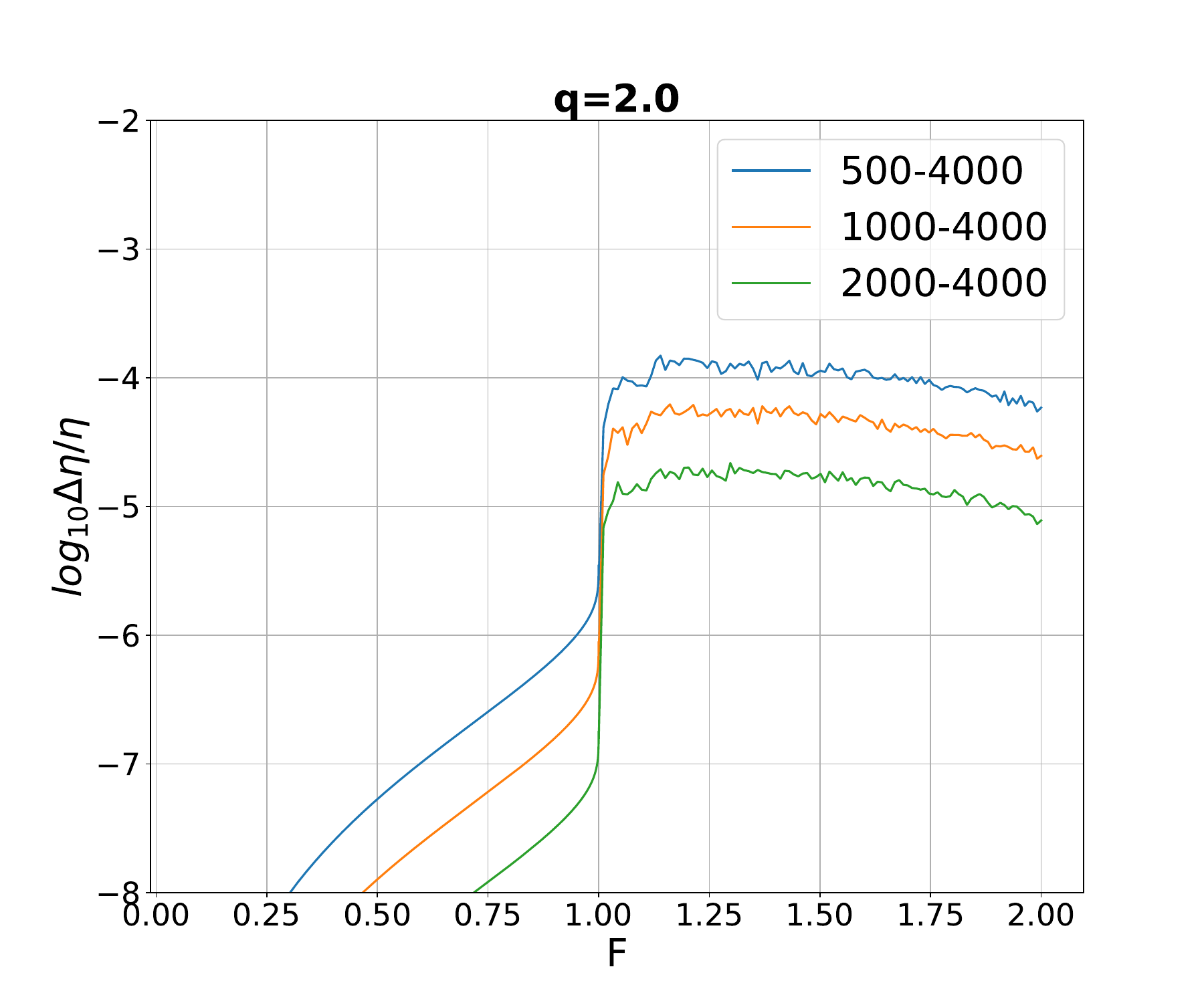}
\caption{ Self-convergence test for the effective acceleration for mass ratios $q=2$. The vertical axis is the normalized error in effective acceleration for angular resolutions $m=500, 1000, 2000$, and $4000$ (Equation \ref{eq:self-converg}). The horizontal axis  is $F$, the fill-out factor $F$ (Equation \ref{eq:fillout}). $F>1$ means the equipotentials are outside of the Roche lobe.}
\label{fig:convergence_acc}
\end{figure}

We have computed the properties of interest for 109 mass ratios in the range of $\log_{10} q \in [-6,5]$. 
We have divided this range into three regimes based on the possible applications it could have:
\begin{eqnarray}
       {\rm \bf A}&:& \ \ \  \log_{10} q\in [-6,-2], \Delta (\log q)=0.25 \label{eq:range1} \nonumber  \\
       {\rm \bf B}&:&  \ \ \ \log_{10} q \in(-2,2], \Delta (\log q)=0.05 \label{eq:range2} \nonumber \\       
       {\rm \bf C}&:&  \ \ \ \log_{10} q \in (2,5], \Delta (\log q)=0.25 \label{eq:range3} \nonumber \\
       \nonumber
\end{eqnarray}
The first range {\bf A} is relevant to the regime of planets or stars orbiting a supermassive black hole, range {\bf B} is a typical range of mass ratios for two stars in a binary system, and range {\bf C} could be used in studies where one wants to investigate the effects of the interactions of a star and a Jupiter-like planet, from the star's perspective. The list of all properties provided for each mass ratio can be found in Appendix A, Table \ref{table:binary_props}. 

We show the volume equivalent radii passing the first three Lagrange points, as a function of the mass ratio in Figure~\ref{fig:lagrangepoint_voleqrad}.

\begin{figure}[ht!]
\includegraphics[width=\columnwidth]{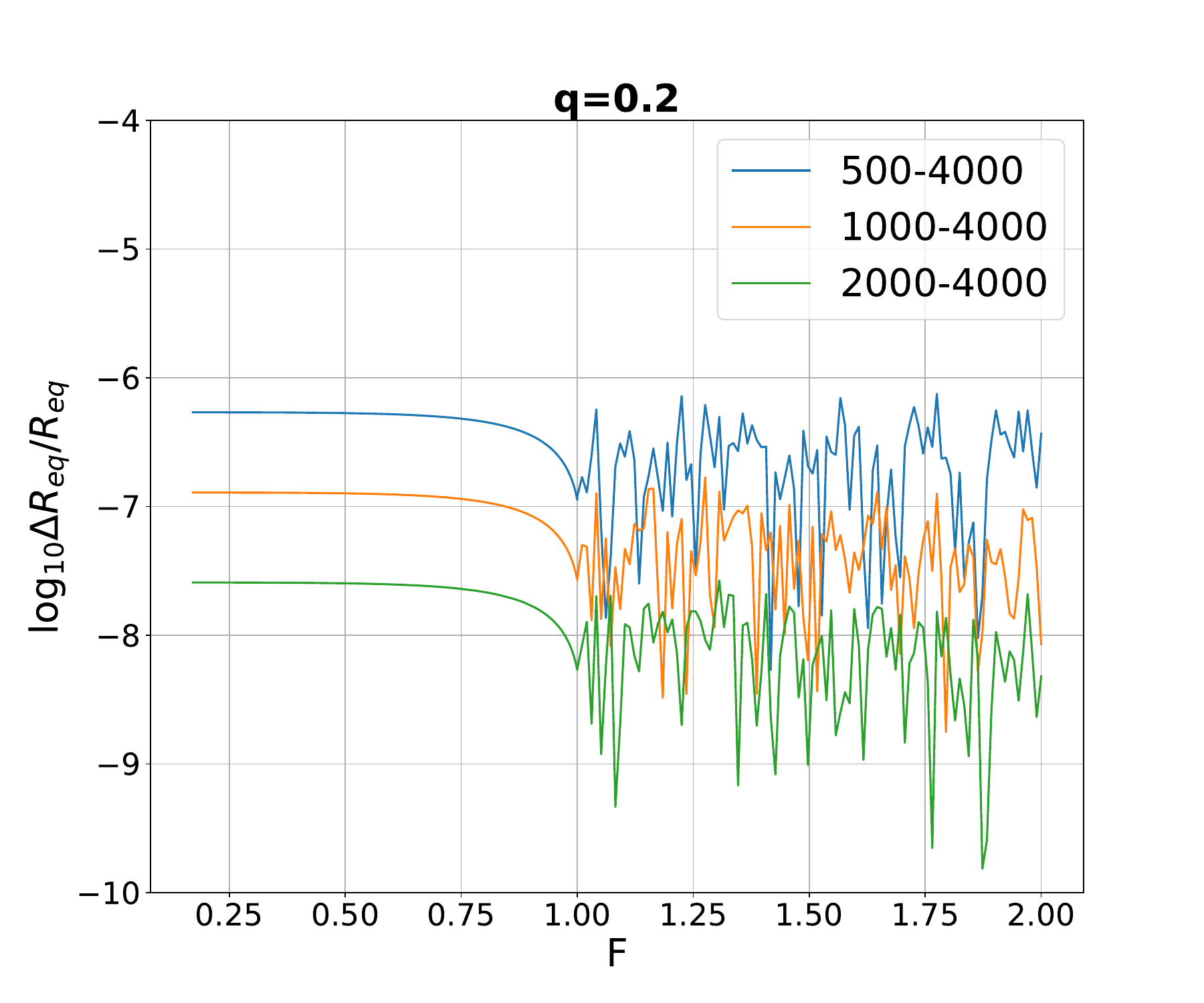}
\caption{ Self-convergence test for the volume-equivalent radius for  mass ratio $q=0.2$. The vertical axis is the normalized error in  volume equivalent radius at $L_1$ for angular resolutions $m=500, 1000, 2000$, and $4000$ (Equation \ref{eq:self-converg}). The horizontal axis is the fill-out factor $F$ (Equation \ref{eq:fillout}).}
\label{fig:convergence_R}
\end{figure}

We investigate at which angular resolution our numerical solver self-converges. Under the self-convergence test, we mean a numerical study in which the resolution of numerical interactions is continuously increased, to see a) if the difference between the numerical outcomes at increasing resolutions continuously decreases, without new features appearing and b) that said difference becomes smaller than the required numerical precision. While self-convergence is often termed as simply convergence, it is important to remember that self-convergence does not verify that the numerical results have converged to a correct solution, for the latter one has to do at least verification tests\footnote{Validation  --  i.e., the measure of numerical model accuracy between model predictions and measurements of the real world -- is not possible in our case, as for most astrophysical problems.}. 
The number of angular zones we considered is $\varphi = 500, 1000, 2000$, and $4000$ for $\varphi$ in the range  $\varphi\in[0,\pi]$ and $\vartheta$ in $\vartheta\in[0,{\pi}/{2}]$.

For a quantity $A$, obtained with the angular resolution $m$, we define as the normalized error

\begin{equation}
    \frac{\Delta A_{m}}{A}=\frac{A_{m}-A_{4000}}{A_{4000}}
    \label{eq:self-converg}
\end{equation}
\noindent We use the highest angular resolution, $n_{\phi}=4000$, to compare the other angular resolutions with it and see whether the deviation is decreasing with increasing angular resolution or not.

In Figures \ref{fig:convergence_acc} and \ref{fig:convergence_R} we provide typical convergences for effective acceleration and volume equivalent radii as functions of the fill-out function $F$. $F$ is a unitless function used to compare different mass ratios' potential distributions up to the outer Lagrange points (which could be $L_2$ or $L_3$), defined as by   \cite{mochnacki1984accurate}:

\begin{eqnarray}
 F&=&\frac{\xi_{L_1}}{\xi}, \; {\rm for}  \;  \xi \geq \xi_{L_1} \ ,  \nonumber  \\
 F&=&1+\frac{(\xi_{L_1}-\xi)}{(\xi_{L_1}-\xi_{L\rm outer})}, \; {\rm for}  \;  \xi \le \xi_{L_1}\ . 
 \label{eq:fillout}
\end{eqnarray}

\noindent $F$ is defined in such a way that $F\in [0,1]$ for potential shells inside and up to the $L_1$ equipotential surface, and $F \in [1,2]$ for shells outside of the $L_1$ equipotential and inside the outer Lagrange point's equipotential shell.
Our results show that self-convergence is obtained quickly. For all mass ratios, the absolute value of the normalized error for effective accelerations is less than $10^{-4}$ ($| \Delta A_m/ A| < 10^{-4} $) for $m = 4000$ outside of the Roche lobe, and is below $5\times 10^{-6}$ within the Roche lobe. Volume-equivalent radii and equipotential areas are obtained with the normalized error less than $10^{-7}$. We, therefore, consider our final results presented in the database and obtained with $n_{\varphi}=4000$ to be self-converged.

\begin{figure}[ht!]
  \includegraphics[width=\columnwidth]{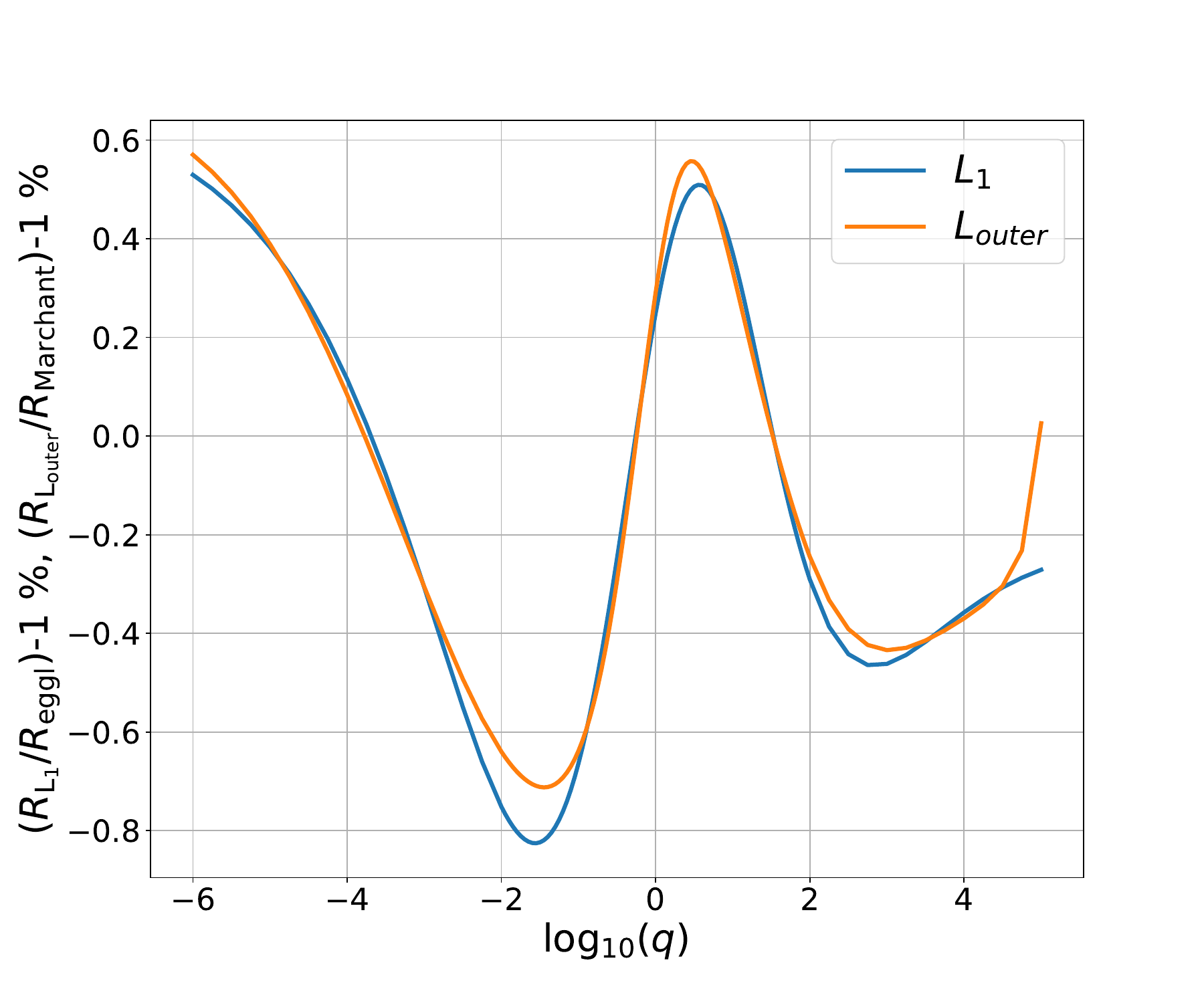}
\caption{Deviation of $R_{L_1}$ ($L_1$ volume-equivalent radius) from \cite{eggleton1983approximations} relation (the blue line). Deviation of $R_{\rm outer}$ (the outer Lagrange equipotential volume-equivalent radius)  from the one provided by the Equation \ref{eq:marchant} (the orange line). The deviations are shown as functions of mass ratio $q$.}
\label{fig:verif_eggleton}
\end{figure}

\label{sec:verifvol}

\subsection{Verification for Volume-Equivalent Radii}

We have verified our volume-equivalent radii against two numerical fits available in the literature (i) the $L_1$ equipotential \citep[][ see also Equation \ref{eq:eggleton} in this paper]{eggleton1983approximations}, and (ii) the outer Lagrange point's equipotential as provided by \citep{marchant2021role}, 

\begin{eqnarray}
        \label{eq:marchant}
        \frac{R_{\rm outer}}{R_{L_1}} &=& 1+\frac{2.74}{1+[(1.02-\ln q)/\sigma]^2} \frac{1}{7.13+q^{0.386}} \ , \nonumber \\ 
        \rm{and} \nonumber \\ 
\sigma&=& \frac{49.4}{12.2+q^{-0.208}} \ . 
\end{eqnarray}

\noindent In the above, $R_{L_1}$ is the radius of the Roche lobe found by using the approximation of \cite{eggleton1983approximations}. Note that here we modified Equation (3) of  \cite{marchant2021role} to match our definition of $q$ (\cite{marchant2021role} define the mass ratio as the companion star's mass divided by the donor's).

The results are shown in Figure~\ref{fig:verif_eggleton}. The fitting equation for $R_{L_1}$ of \citet{eggleton1983approximations} has been reported (in the paper where the equation was presented) to be accurate to $1\%$. The fitting equation for $R_{\rm outer}$ by \cite{marchant2021role} has been reported by them to be accurate to 0.15\%. However, the equation  for $R_{\rm outer}$ also, by definition, includes the uncertainty in $R_{L_1}$ from the formula of \citet{eggleton1983approximations}. We find that  we have agreement within $1\%$ for both fitting equations.

\begin{figure*}[t!]
\centering
\includegraphics[width=\columnwidth]{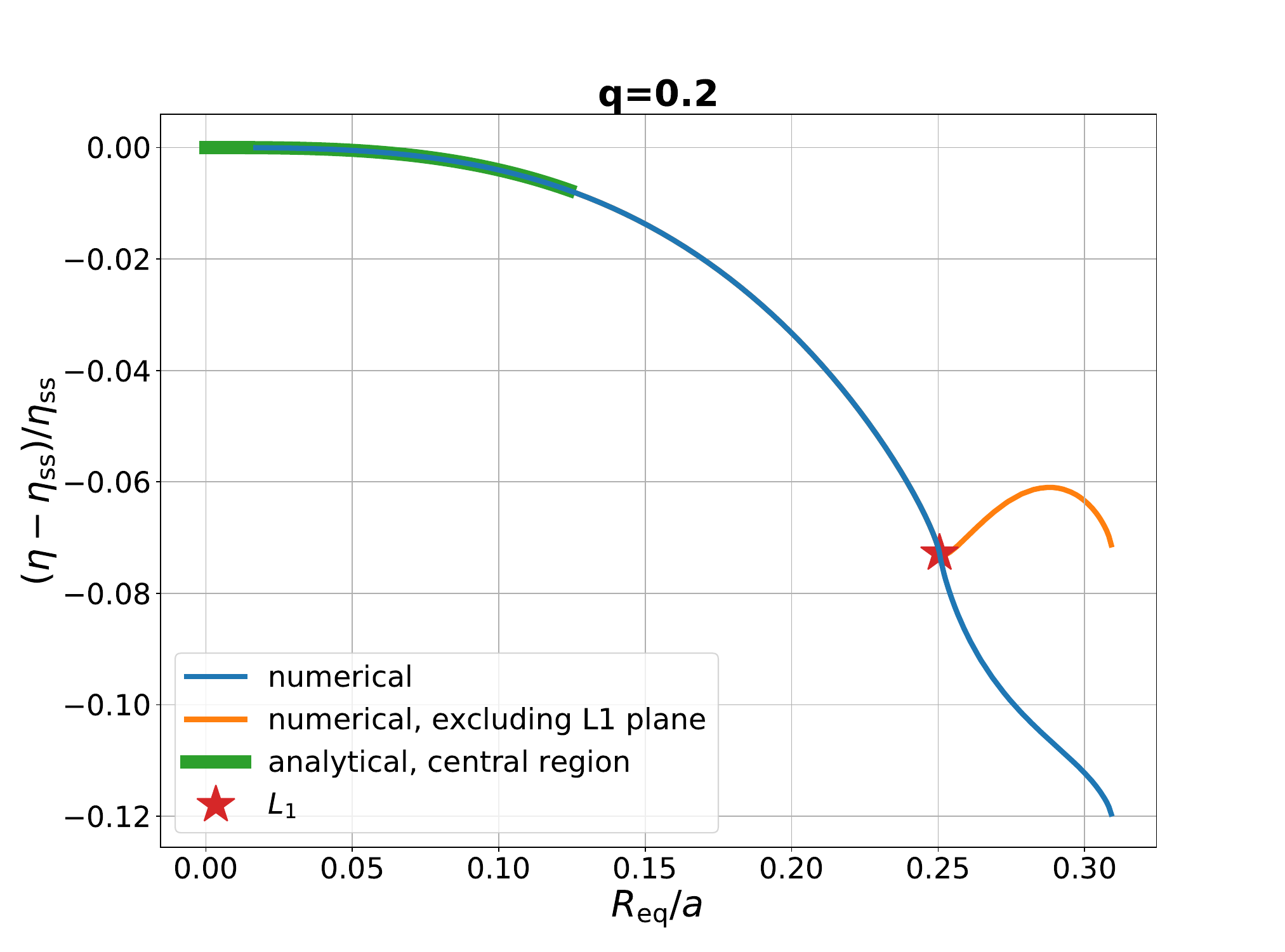}
\includegraphics[width=\columnwidth]{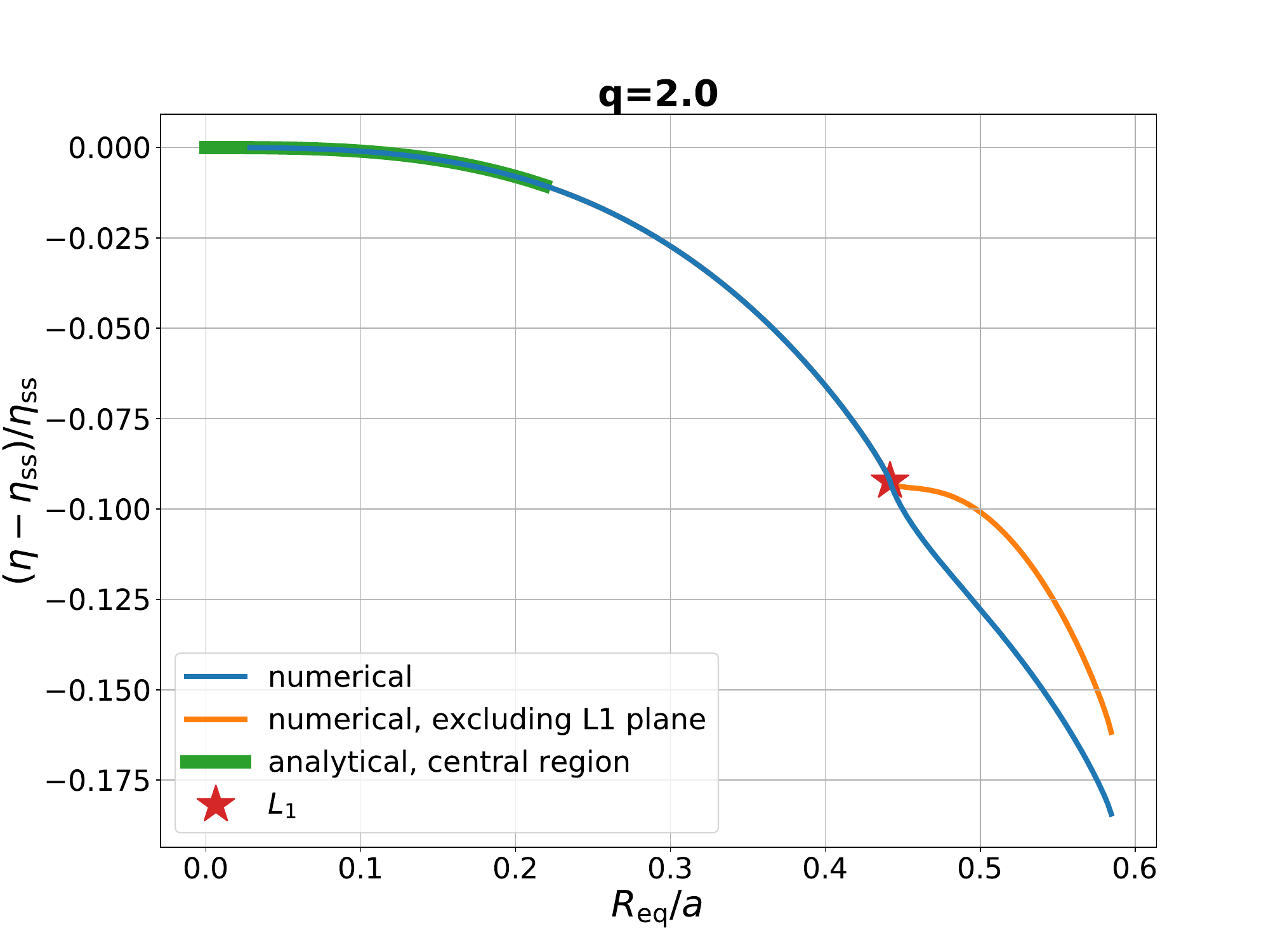}
\caption{The deviation in effective acceleration from our tables for  mass ratios $q=0.2$ (the left plane) and $q=2$ (the right plane), as a function of the volume equivalent radius. The red star denotes the equipotential passing $L_1$. The green curve in each figure is the analytical expression \ref{eq:deviation_close_tocenter}. We show the analytical solution for the distances within $0.5$ of $R_{L1}$, albeit we use it only for the distances within $0.05$ of $x_{L_1}$, as described in the text. One can see that the analytical solution matches the numerical solution for a large range of distances. The blue line shows the case when the effective acceleration is averaged over the whole contour surrounding the star, and the orange line shows the effective acceleration with $L_1$ plane is excluded from the averaging.}
\label{fig:deviation}
\end{figure*}

\subsection{Verification for Effective Acceleration}

\cite{mochnacki1984accurate} has provided coarse tables with numerically obtained average effective acceleration on the equipotential as a function of the potential.
We have compared our average effective acceleration passing the $L_1$ shell and the $L_2$ shell (obtained with $n_{\varphi}=4000$) with that of \citet{mochnacki1984accurate}. 
The deviations are within $10^{-4}$ for the values at the outer Lagrangian equipotential (their $F=2$), and within the last significant digit of the values that \citet{mochnacki1984accurate} provides for $L_1$ equipotential (their $F=1$). These deviations are consistent with our test for self-convergence. 

\subsection{Deviation of the binary effective acceleration from spherically symmetric gravitational acceleration}

\label{sec:overall}

The final result of the effective acceleration is shown 
in Figure~\ref{fig:deviation}. The analytical solution at distances close to the center has also been plotted and, as one can see, the numerical results correspond to the analytical equation very closely.

\begin{figure}[ht!]
\centering
\includegraphics[width=\columnwidth]{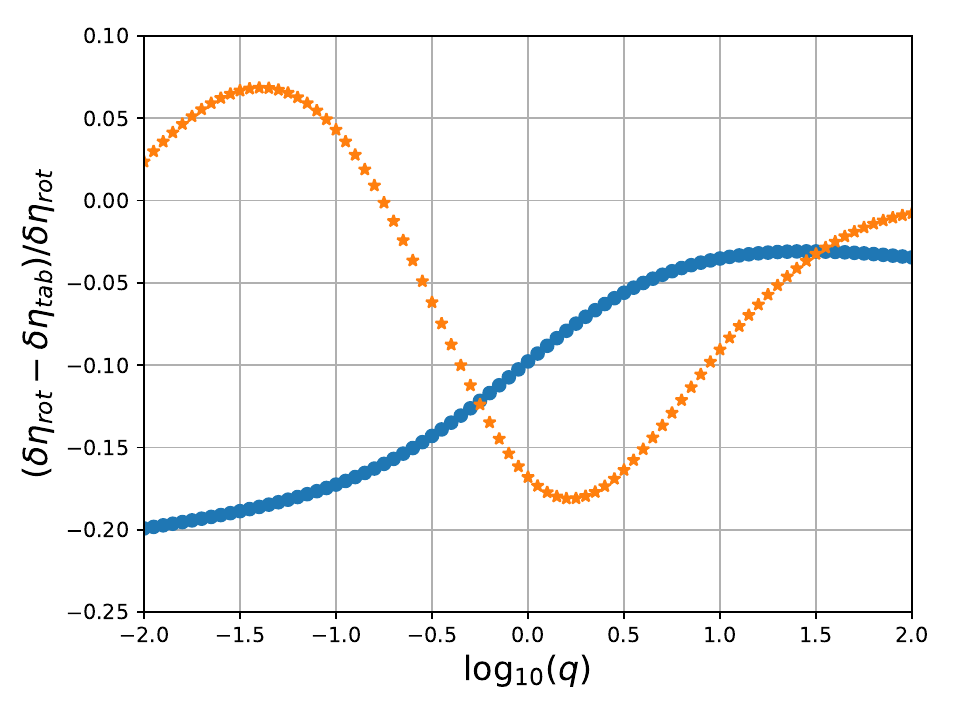}
\caption{The deviation in effective acceleration from our tables from a solid body rotating star, as a function of the mass ratios (blue circles). The deviation is provided for stars that fill their Roche lobe. With orange stars, we show the deviation where we used Eggleton's equation for Roche lobe radius instead of our integrated quantities for the spherically symmetric acceleration and the rotating term. See \S\ref{sec:rotating}. }
\label{fig:deviation_sph}
\end{figure}

\begin{figure}[ht!]
\centering
\includegraphics[width=\columnwidth]{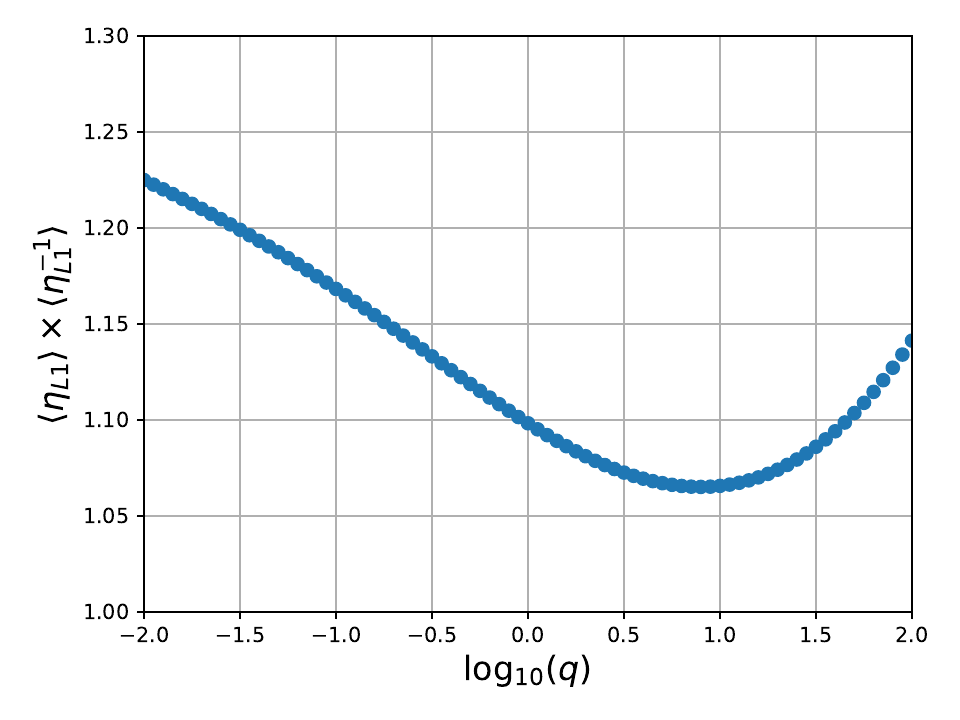}
\caption{The effective gravity averaged over equipotential multiplied by the inverse effective gravity averaged over equipotential, as a function of the mass ratios. See \S\ref{sec:inverse}.}
\label{fig:gginv}
\end{figure}

\begin{figure}[ht!]
\centering
\includegraphics[width=\columnwidth]{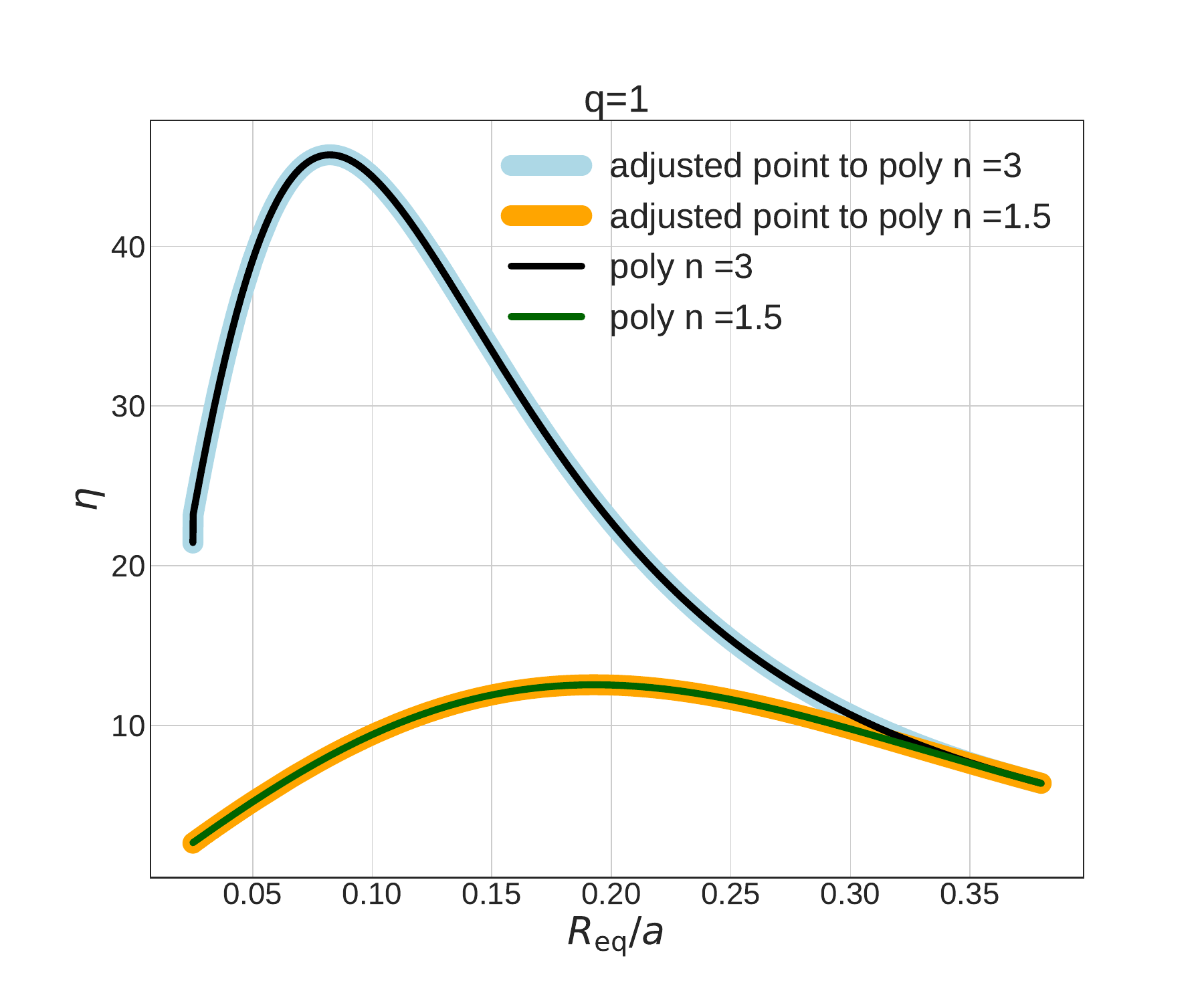}
\caption{ The simplified effective acceleration and the integrated effective acceleration, as a function of $R_{\rm eq}$, for polytropes  with $n=3$ (blue line) and $n=1.5$ (orange line). Shown for the case $q=1$. }
\label{fig:polytropes}
\end{figure}

\subsection{Deviation of the binary effective acceleration from the case of a solid body rotating star}

\label{sec:rotating}

If a star is a solid body rotator, it experiences an additional acceleration that opposes gravitational acceleration. If one assumes spherical symmetry, the effective acceleration, averaged over a spherical shell, can be expressed in the units of acceleration in a non-rotating star as

\begin{equation}
\frac{\eta_{\rm rot}}{\eta_{\rm ss}} = 1 -\frac{2}{3} \frac{1+q}{q} r^3 =  1+ \delta \eta_{\rm rot} \ .
\end{equation}

\noindent Here we introduced the relative deviation that rotation causes, $\delta \eta_{\rm rot}$. Analogously, one can consider that our numerically obtained and tabulated acceleration is

\begin{equation}
\frac{\eta}{\eta_{\rm ss}} =  1+ \delta \eta_{\rm tab} \ .
\end{equation}

In Figure~\ref{fig:deviation_sph}, we compare those two terms for the binaries with the mass ratios range in the interests of stellar binaries. The comparison is provided for stars presumed to fill their Roche lobe -- their radii are taken to be equal to their volume-equivalent Roche lobe radii. The effective binary potential always results in the average effective acceleration being smaller  (in its absolute value) than in the case of a spherically symmetric star that rotates with an angular velocity equal to the binary angular velocity.

\subsection{Alternative treatment of the effective acceleration}

\label{sec:inverse}

It has been argued that instead of the effective gravity $g$ (averaged over the equipotential), the quantity that should be averaged on the equipotentials for the inclusion in the stellar equations is $\langle g^{-1} \rangle$, \citep[see][for how to use it]{KT70,ES76, ES78}. We provide the unitless quantity $\langle g^{-1} \rangle$ in our numerical tables. The quantity $\langle g^{-1} \rangle ^{-1}$ deviates from spherically symmetric gravity even more significant than $\langle g  \rangle$, as can be seen from Figure~\ref{fig:gginv} where we show the values on $L_1$ equipotential for a range of mass ratios $ \langle g \rangle \times \langle g^{-1} \rangle = \langle\eta \rangle \times \langle \eta^{-1} \rangle$.

\subsection{How to use the provided quantities}

\label{sec:howto}
 
This work is devoted to the evolution of stars within the gravity field of a binary system. Hence, throughout this paper, we use the same definition of the effective acceleration as per Equation~\ref{eq:tot_eta} -- we always consider the magnitude of the effective acceleration vector component that is normal to the equipotential surface at each considered location. The adopted definition of the effective acceleration holds for values obtained on the $ L_1$ plane, including the inverse values of the effective acceleration. In the main data table (see Appendix \ref{sec:tabdescr} for more detail), we provide:

\begin{enumerate}[(a)] 
\item values from 3D integrations made throughout the closed 3D surface (including the $L_1$-plane)
\item values from 2D integrations for the $L_1$ plane only.
\end{enumerate}

When the considered star overfills its Roche lobe, the values on the truncated (limited by the $L_1$-plane) equipotential layers of the star are to be reconstructed from the 3D values provided in a), excluding the 2D values provided in b). For example, the effective acceleration at $i$-equipotential shell inside the star is to be found using the 3D area $A$, the averaged (over $A$) 3D effective acceleration $\eta$, the 2D area of $L_1$-plane $A_{\rm Lpl}$, and the averaged (over the $L_1$-plane) effective acceleration $\eta_{\rm Lpl}$:

\begin{equation}
\eta_i  = \frac{\eta A - \eta_{\rm Lpl} A_{\rm Lpl}}{A - A_{\rm Lpl}} \ .
\end{equation}

We also provide the auxiliary data table with values from 3D integrations made throughout the truncated surface only, excluding the $L_1$-plane. This table can be used directly without reconstruction.
Note that the angle-distributed mesh does not always fall on the cross-section of the 3D equipotential lines and the $L_1$-plane.
Hence, the truncated 3D area will not have a continuous transition to the $L_1$-plane. This 3D integration is prone to carry errors due to an incomplete layer of triangular mesh between 3D equipotential and flat $L_1$-plane.
While both methods provide essentially the same value, the second has a more significant numerical convergence error, and we strongly recommend using the reconstruction method.

For the $L_1$-plane, the quantities in the main data table are not related to the component of the effective acceleration that is perpendicular to the $L_1$-plane, $\eta_{x,L1}$. Suppose one needs values for the $x-$component of the acceleration only, that can be easily obtained doing 2D integration on the $L_1$-plane for $\eta_x$ with $x=x_{L1}$ (see Equation~\ref{eq:etax}), and then recovering the value with physical units by multiplying it by $-G(M_1+M_2)/2a^2$. While this quantity is not the primary goal of this work, we supplement the averaged values for the $\eta_x$-component separately in another auxiliary data table (see Appendix \ref{sec:tabdescr}).

\section{Effects of a non-point mass}

\label{sec:nonpoint}
A real star is not a point mass. One has to obtain a 3D stellar model to include the effect of a realistic 3D mass distribution on the effective acceleration. Here,  we only consider a first-order effect. We start with defining the local mass fraction as a function of the distance to the center of the first star

\begin{equation}
    q_{\rm loc} (r_{\rm eq}) \equiv\frac{M_{\rm loc}(r_{\rm eq})}{M_2} \ ,
\end{equation}

\noindent where $q_{\rm loc}(r_{\rm eq})$ is the mass fraction that assumes the mass within the volume-equivalent radius, $M_{\rm loc}(r_{\rm eq})$ is the local mass coordinate inside the donor, $M_{\rm loc}(r_{\rm eq})\le M_1$.

We start by considering the deviation in the gravitational  potential due to a non-point mass distribution inside the first star, keeping the second star's gravitational contribution, and the effect from being in the corotating frame, the same as previously

\begin{eqnarray}
\psi(\vec R) &=& - G \int_{M1} \frac{dm}{|R_1|} 
 -\frac{GM_2}{|R_2|} \nonumber \\
 & & -\frac{1}{2}\Omega^2\left [\left (x-\frac{a M_2}{M_1+M_2}\right )^2+y^2\right ]
  \label{eq:localpot}
\end{eqnarray}

\noindent Here, the first term is, in principle, a 3D integral that has to be taken through the donor star's interior.
The effective acceleration is the gradient of the effective potential. We consider the changes in the first term of the effective potential due to the mass having a spherically symmetric but non-point mass distribution. The former assumption takes us to the Newton shell theorem, which states that the gravitational field due to the mass inside a given radius acts as a point mass to elements at that radius, whereas the mass outside of that radius does not exert a net gravitational force 
on those elements.
In the case of spherically-symmetric mass distributions, the unitless effective acceleration can be written then as
\begin{eqnarray}
       \eta_{x,\rm loc}(x,y,z)&=&\frac{2xq_{\rm loc}(r_{\rm eq})}{(1+q)|r_1|^3}
       +\frac{2(x-1)}{(1+q)|r_2|^3}  \nonumber \\ \nonumber 
       & - & 2(x-\frac{1}{1+q}) \ , \\ \nonumber 
     \eta_{y,\rm loc}(x,y,z)&=&
     \frac{2yq_{\rm loc}(r_{\rm eq})}{(1+q)|r_1|^3}+\frac{2y}{(1+q)|r_2|^3}-2y \ , \nonumber \\
     \eta_{z,\rm loc}(x,y,z)&=&
     \frac{2zq_{\rm loc}(r_{\rm eq})}{(1+q)|r_1|^3}+\frac{2z}{(1+q)|r_2|^3} \ .
     \label{eq:local_acc}
\end{eqnarray}
\noindent The total acceleration at the given location can be found as 

\begin{equation}
\eta_{\rm loc}(r)=\sqrt{\eta_{x,\rm loc}^2+\eta_{y,\rm loc}^2+\eta_{z,\rm loc}^2}\ .
\end{equation}

Based on the shape of the local potential, we consider the correctional term for the effective acceleration as 

\begin{equation}
\delta \eta \left (q_{\rm loc}(r_{\rm eq}),q,r_{\rm eq}\right ) =  \frac{2 \left ( q_{\rm loc}
        (r_{\rm eq})-q\right )}{(1+q)} \frac{1}{r_1^2} \ .
\label{eq:deta}
\end{equation}

\noindent We also introduce a simple ''modified''
effective acceleration as 

\begin{equation}
\eta_{\rm mod}(q_{\rm loc}(r_{\rm eq}),q,r_{\rm eq})=\eta(q,r_{\rm eq}) + \delta \eta(q_{\rm loc}(r_{\rm eq}),q,r_{\rm eq}) \ . 
\label{eq:modified_acc}
\end{equation}

\noindent  Here, $\eta(q,r_{\rm eq})$ is the same unitless effective acceleration as previously, for point-mass binaries, and its values for each $q$ and $r_{\rm eq}$ can be found from the tables we obtained and provided for the reader. 

In order to test this approximation, we considered polytropes with $n=1.5$ and $n=3$. For the local mass ratio, we assumed that the mass within each volume-equivalent of the numerically obtained binary equipotential (each equipotential itself is a function of the local mass) is the same as within the same volume in a spherically symmetric polytrope.
As shown in Figure~\ref{fig:polytropes}, the exact integration and the simplified method are in agreement with each other. We conclude that the simplified method, as described, can be used to obtain effective accelerations inside the star from our precalculated tables for point mass cases and the correction term. 
It should be noted that, for regions close to the center, $r_1 \leq 0.05x_{L_1}$, we make use of the following analytical expression, to obtain the effective acceleration:

\begin{equation}
\eta = \eta_{\rm ss,loc} + \Delta \eta_{\rm loc} ; \quad 
{\Delta \eta_{\rm loc}} = - \frac{2}{3} \frac{ (1+q)}{q_{\rm loc}(r_1)}  r_1^3 \eta_{\rm ss,loc}.
\label{eq:dev_close_nonpoint}
\end{equation}

\noindent It is assumed that, in the vicinity of the center, $r_{\rm eq}\simeq r_1$. In this equation, $\eta_{\rm ss,loc}$ is the effective acceleration of a single, non-point mass star up to that zone:\\
\begin{equation}
\eta_{\rm ss,loc} =  \frac{2q_{\rm loc}(r_1)}{(1+q)} \frac{1}{r_1^2} .
    \label{eq:ss_np}
\end{equation}
The derivation of these equations is written in \ref{app:eff_acc_np}.

The subroutine that we provide (see the description in Appendix \ref{ap:sub}) makes use of the described method.
\section{Conclusions}

\label{sec:conclusion}

We have described the method we use to obtain numerically effective accelerations for a star in a binary system as a function of either the effective potential, or the volume-equivalent radius.

Our method is inspired by the fact that isobaric surfaces coincide with the equipotential shells when stars are in the hydrostatic equilibrium. This makes the concepts of volume-effective radii and average effective acceleration of each shell a valuable tool to simulate binary stars in a 1D stellar evolution code. 

Our method for obtaining the volume-equivalent radii of each equipotential shell is to integrate with spherical volume elements from the vicinity of the center of the donor star up to the point where the corresponding potential is equal to that sought, with fixed precision. 
The average effective acceleration on each shell is obtained by dividing each shell into triangular mesh areas and then calculating the weighted average of the effective acceleration, with the weight being the area of the triangular mesh element. The integration area includes  $L_1$-plane. In addition, we obtained the effective accelerations averaged on the $L_1$-plane as a function of the cross-sectional area or effective potential (the semi-major and semi-minor axes of these ellipse-like cross-sections are also tabulated).   We also obtained the average inverse effective acceleration, averaged over the entire equipotential level and on the $L_1$-plane. Values on the equipotential shell only (excluding $L_1$-plane) can be recovered from the two provided values, see \S~\ref{sec:howto}.

We provide the tables for a range of mass ratios, from $10^{-6}$ to $10^5$.  We have verified the provided tables for self-convergence by taking the run for several angular resolutions.
We have also validated the tables by comparing our results to the fit provided for the volume equivalent radius of the shell passing through $L_1$ by \cite{eggleton1983approximations}, the fit provided for the volume equivalent radius of the shell passing through the outer Lagrange point by \cite{marchant2021role} and  the average effective acceleration of shells passing through $L_1$ and $L_2$ tabulated by  \cite{mochnacki1984accurate} for certain mass ratios.

Further, we provide analytical expressions describing the effective acceleration's behavior in the vicinity of the center of the donor star in order to check whether our numerical tables work well close to the center. We also describe a rapid method to obtain effective accelerations for the non-point mass profiles that can take place in stars,  using our pre-calculated table for point-mass calculations. This rapid method has been tested and verified to agree with the results for integrating the polytropic mass distribution for $n=1.5$ and $n=3$, encompassing most of the mass distributions that are plausible in stars.

We have provided a subroutine for the community that can be used in any 1D stellar code to rapidly obtain effective accelerations (as a function of the local mass and radius) that are more appropriate for the case of binary stars.
We hope that the method, the tables, and the subroutine will help with further progress in the understanding of binary star evolution.

\begin{acknowledgments}
N.I. acknowledges funding from NSERC Discovery under Grant No. NSERC RGPIN-2019-04277. 
This research was enabled in part by support provided by Compute Canada (\url{www.computecanada.ca}). We thank the referee for the valuable suggestions that helped to improve the manuscript.
\end{acknowledgments}

\vspace{5mm}
\facilities{Compute Canada}

\appendix

\section{Description of the tables}
\label{sec:tabdescr} 

We provide one file for each of the 109 calculated mass ratios $q$ between $10^{-6}$ and $10^5$. The file's name contains the value of $\log_{10} q$ for which it was obtained. Each file has 15 columns of data, as described in Table \ref{table:binary_props}. 
There are 600 data points (rows) for each table, starting from $\xi=\xi(0.05x_{\rm L_1},0,0)$ to $\xi=\xi_{\rm Louter}$. The equipotential shell passing via $L_1$ is the data point number 500.

Columns 3--8 of data provide values at the equipotential shells at all data points as obtained from 3D integrations.
Hence, the equipotential shells are inclusive of the $L_1$-plane; the truncated equipotential areas and values
on them are to be reconstructed as described in \S~\ref{sec:howto}.
The columns 9--14 provide data for the $L_1$-plane 
but only for mass ratios $\log_{10} q \leq 2.5$ (a regime where stellar interactions may lead to a mass transfer with a non-negligible mass stream thickness).
The $L_1$-plane properties are given only at the rows 501-600 for the equipotentials less to the $L_1$ value. In rows 1--500, columns 9--14 are filled with the ``0'' value as a placeholder. 
For $\log_{10} q > 2.5$,  columns 9-14 for rows 501-600 are filled with the ``0'' value. For point mass cases, one needs three quantities, the donor star mass $M_1$, the mass ratio $q$, and the binary separation $a$, to recover values in CGS. The conversions can be found in Table \ref{table:binary_props}.

Another set of files we provide on the website includes a compact version of the previously mentioned. In this version, the units of potential and effective acceleration have been provided in terms of a single point-mass star's potential and acceleration, respectively. This version of the table and the relevant conversion units are tabulated in Table \ref{table:comp_binary_props}. This table is used by the code described in Appendix \ref{ap:sub} to obtain effective accelerations. 

In addition, a file has been provided with properties corresponding to the first three Lagrange points for all mass ratios computed, see Table \ref{table:rL_1l2l3}. For mass ratios lower than 1, where the outer Lagrange point is $L_2$, and $L_3$ is on the opposite side, we have filled $-1$ for properties corresponding to $L_3$; the same has been applied for columns 5 and 6 of mass ratios with $\log_{10} q > 2.5$.

There are two more supplementary tables (see \S\ref{sec:howto}). One of them is very similar to Table \ref{table:binary_props}, but 3D quantities were obtained while doing integrations over truncated equipotentials. The second supplementary table provides values of $\eta_x$ integrated on $L_1$-plane.

\begin{deluxetable}{llll}

\tabletypesize{\scriptsize}

\tablecaption{The properties computed and included in our database for each mass ratio.}

\label{table:binary_props}

\tablehead{\colhead{}  & \colhead{quantity} & \colhead{notation} & \colhead{conversion equation} \\ 
\colhead{} & \colhead{} & \colhead{} & \colhead{} } 

\startdata
1 & equipotential shell number \\  
2 & mass ratio &  $q$ &  unitless \\
3 & fill-out function &  F &  unitless \\
4 & potential & $\xi$& $\xi \times \frac{-GM_1(1+q)}{2qa}$ \\
5 & volume-equivalent radius &  $R_{\rm eq}$ &  $R_{\rm eq} \times a$ \\
6 & effective acceleration averaged on the equipotential shell &  $\eta_{\rm shell}$ &  $\eta_{\rm shell} \times \frac{GM_1(1+q)}{2qa^2}$ \\
7 & inverse effective acceleration averaged on the equipotential shell &  $\langle \eta_{\rm shell}^{-1} \rangle $ &  $\langle \eta_{\rm shell}^{-1} \rangle  \times \frac{2qa^2}{GM_1(1+q)}$ \\
8 & area of the equipotential shell &  $A$ &  $A \times a^2 $\\
9 & area of $L_1$ cross-section &$A_{\rm Lpl}$ & $A_{\rm Lpl}\times a^2$ \\
10 & y intersection of $L_1$ cross-section &$y_{\rm Lpl}$ & $y_{\rm Lpl}\times a$\\
11 & z intersection of $L_1$ cross-section & $z_{\rm Lpl}$ & $z_{\rm Lpl}\times a$ \\
12 & effective acceleration averaged over the intersection with $L_1-$plane &  $\eta_{\rm L}$ &  $\eta_{\rm L} \times \frac{GM_1(1+q)}{2qa^2}$ \\
13 & effective acceleration averaged over $L_1$ cross-section area &  $\eta_{\rm Lpl}$ &  $\eta_{\rm Lpl} \times \frac{GM_1(1+q)}{2qa^2}$ \\
14 & inverse effective acceleration averaged  over $L_1$ cross-section area &  $\langle \eta_{\rm Lpl}^{-1}\rangle $ &  $\langle \eta_{\rm Lpl}^{-1}\rangle  \times \frac{2qa^2}{GM_1(1+q)}$ \\\enddata

\end{deluxetable}

\begin{deluxetable}{llll}

\tabletypesize{\scriptsize}

\tablecaption{The compact version of the database that will be used in \ref{ap:sub}}

\label{table:comp_binary_props}

\tablehead{\colhead{} & \colhead{quantity} & \colhead{notation} & \colhead{conversion equation} \\ 
\colhead{} & \colhead{} & \colhead{} } 

\startdata
1 & volume-equivalent radius &  $R_{\rm eq}$ &  $R_{\rm eq} \times a$ \\
2 & relative potential & $\xi_{\rm rel}$& $\xi_{\rm rel} \times \frac{-GM_1}{r_{\rm eq}}$ \\
 3 & relative average effective acceleration on equipotential shell &  $\eta_{\rm rel,shell}$ &  $\eta_{\rm rel,shell} \times \frac{GM_1}{r_{\rm eq}^2}$ \\
4 & relative average effective acceleration on $L_1$ cross-section &  $\eta_{\rm rel,Lpl}$ &  $\eta_{\rm rel,Lpl} \times \frac{GM_1}{r_{eq}^2}$ \\
5 & area of $L_1$ cross-section &$A_{\rm Lpl}$ & $A_{\rm Lpl}\times a^2$ \\
6 & area of equipotential shell  &$A$ & $A\times a^2$ \\
\enddata

\end{deluxetable}

\begin{deluxetable}{llll}

\tabletypesize{\scriptsize}

\tablecaption{The properties computed and included in our database for each of the first three Lagrange points.}

\label{table:rL_1l2l3}

\tablehead{\colhead{} & \colhead{quantity} & \colhead{notation} & \colhead{conversion equation} \\ 
\colhead{} & \colhead{} & \colhead{} } 

\startdata
1 & mass ratio &  $q$ &  unitless \\
2 & volume-equivalent radius of shell passing $L_1$ &  $R_{\rm eq,L1}$ &  $R_{\rm eq,L1} \times a$ \\
3 & volume-equivalent radius of shell passing $L_2$ &  $R_{\rm eq,L2}$ &  $R_{\rm eq,L2} \times a$ \\
4 & volume-equivalent radius of shell passing $L_3$ & $R_{\rm eq,L3}$ & $R_{\rm eq,L3} \times a$ \\
5 & area of $L_1$ cross-section of $L_2$ shell &  $A_{\rm Lpl,L2}$ & $A_{\rm Lpl,L2}\times a^2$\\
6 & area of $L_1$ cross-section of $L_3$ shell &  $A_{\rm Lpl,L3}$ & $A_{\rm Lpl,L3}\times a^2$\\
7 & average effective acceleration on the L-plane of $L_2$ shell &  $\eta_{\rm Lpl,L2}$ &  $\eta_{\rm Lpl,L2} \times \frac{GM_1(1+q)}{2qa^2}$ \\
8 & average effective acceleration on the L-plane of $L_3$ shell &   $\eta_{\rm Lpl,L3}$ &  $\eta_{\rm Lpl,L3} \times \frac{GM_1(1+q)}{2qa^2}$ \\
\enddata

\end{deluxetable}

\section{subroutine for effective acceleration}

\label{ap:sub}

The subroutine that we provide uses as input $M_1$,  $r_{1}$ (distance from the center of the donor star), $M_{\rm 1, loc}$ (enclosed mass of the star from the center up to $r_{1}$), $M_2$, and $a$. By choice of the user, it returns either value of the effective acceleration in CGS, or the effective acceleration as a fraction of the single star's effective acceleration. The values are for the truncated equipotentials.
Our subroutine that uses these databases  gives the effective acceleration as a function of the masses of the two stars, the orbital separation, and the distance from the center of the donor star. 
Once the mentioned inputs to the subroutine are given, the code determines between which two mass ratios does the local mass ratio fall in between. Then for those two tables it finds where the distance from the donor star's center falls in between and interpolates for the effective acceleration between the two volume equivalent radii. After doing this for both mass ratios, it interpolates again, this time between mass ratios and the two obtained accelerations corresponding to those mass ratios to obtain the finalized effective acceleration. This acceleration could be given in CGS units, or as a unitless parameter by dividing it by the local gravitational acceleration if there was only a single star.

If the mass ratio specified is lower than the minimum mass ratio in the database ($10^{-6}$), or larger than the maximum mass ratio ($10^{+5}$), the code will output an error. If the distance and orbital separation is specified such that $r_{1}/a \le 0.05$, the subroutine will use the analytical expression \ref{eq:deviation_close_tocenter} to obtain the effective acceleration instead. The code as it is should not be used for the cases when the mass ratio is above $10^{+2.5}$, and there is RLOF. In the latter case, the user should re-do this table using 3D quantities obtained while doing integrations over truncated equipotentials.
The databases and sample subroutines are available at Zenodo \citep{ZenodoTables}, and updates will be  available at \url{https://github.com/AliPourmand/1D_binary_star_properties}.

\section{Behavior close to the center of $M_1$.}
\label{app:near_center}

We consider a coordinate system in corotation with the binary, with the origin at the donor star's center. The $xy$ plane is the plane of the orbit, and the co-rotating plane is rotating around the axis which passes through the center of mass of the binary system. 
The effective potential $\Psi$ in a spherical coordinate system that rotates with the fixed binary orbital angular velocity  $\Omega=\sqrt{G(M_1+M_2)/a^3}$:

\begin{eqnarray}
    \Psi (R,\vartheta,\varphi) &=& -\frac{GM_1}{R}-\frac{GM_2}{\sqrt{(R\cos \varphi \sin \vartheta-a)^2+R^2\sin^2 \varphi \sin^2 \vartheta}+R^2 \cos^2 \vartheta} \nonumber \\
     & & -\frac{1}{2}\Omega^2\left [(R\cos \varphi \sin \vartheta-a\frac{M_2}{M_1+M_2})^2+R^2\sin^2 \varphi \sin^2 \vartheta )\right ] \ .
     \label{eq:potential_app}
\end{eqnarray}

\noindent Here  $M_1$ and $M_2$ are the donor and companion star masses, $a$ is the orbital separation, and $R$ is the distance to the origin.

The first term $\Psi_{\rm ss}$ is the potential of a spherically symmetric single star of mass $M_1$. The other two terms represent the deviation of the effective potential $\Delta \Psi=\Psi_{\rm c}+\Psi_{\rm b}$ due to having a companion of mass $M_2$, $\Psi_{\rm c}$, and due to being in a coordinate system that rotates with $\Omega$, $\Psi_{\rm b}$. Introducing $r=R/a$, the terms are reduced to:

\begin{eqnarray}
    \Psi_{\rm ss} (r,\vartheta,\varphi) &=&
    -\frac{GM_1}{a}\frac{1}{r} \ , \nonumber \\
    \Psi_{\rm c} (r,\vartheta,\varphi)&=&  -\frac{GM_2}{a} \frac{1}{\sqrt{r^2 - 2r \cos \varphi \sin \vartheta+1}} \ ,
    \nonumber \\ 
     \Psi_{\rm b} (r,\vartheta,\varphi) & =& -\frac{1}{2}\frac{G{(M_1+M_2)}}{a} \left [
     r^2 \sin^2 \vartheta - 2r\frac{M_2}{M_1+M_2} \cos \varphi \sin \vartheta + \frac{M_2^2}{(M_1+M_2)^2}\right ]   \ .
\end{eqnarray}

We want to investigate the behavior of the effective acceleration and potential at distances close to the center; thus we assume that equipotential surfaces there are almost spherical (we have verified this assumption by numerical integrations).

\subsection{Effective acceleration}

\label{app:eff_acc}

As the equipotential surfaces are almost spherical,  we adopt that for the effective acceleration, the only derivative of the effective potential that matters is the derivative by $r$. 

\
\begin{eqnarray}
 \frac{\partial  \Delta \Psi (r,\vartheta,\varphi) }{\partial r} = \frac{G}{a} \left [ M_2 \frac{(r-\cos \varphi \sin \vartheta)}{(r^2 - 2r \cos \varphi \sin \vartheta+1)^{3/2}} 
- (M_1+M_2)  r \sin^2 \vartheta   +M_2  \cos \varphi \sin \vartheta  \right ] \ .
\end{eqnarray}

In the limit $r\ll L_1$, 

\begin{eqnarray}
& &\left . \frac{\partial  \Delta \Psi (r,\vartheta,\varphi)}{\partial r} \right |_{r\ll 1} =   \frac{G}{a} \left [
M_2  (r  - 3r \cos^2 \varphi \sin^2 \vartheta )
 - (M_1+M_2)    r\sin^2 \vartheta   \right ] \ .
\end{eqnarray}

After the integration over all angles, the only contributing term that remains is from being in the co-rotating frame, while the contribution of the gravitational field of the companion has vanished:

\begin{eqnarray}
& & \left \langle \left . \frac{\partial  \Delta \Psi (r,\vartheta,\varphi)}{\partial r}\right |_{r\ll 1} \right \rangle = \frac{1}{4 \pi} \int_{4 \pi} \left . \frac{\partial \Delta \Psi (r,\vartheta,\varphi)}{ \partial r} \right |_{r\ll 1} d\Omega =   - \frac{2}{3} \frac{G (M_1+M_2)}{a}  r \ .
\end{eqnarray}

 Finally, the relative deviation of the gradient of the effective potential in a binary, from the case of the spherically symmetric one, near the center, is 

\begin{eqnarray}
& & \left \langle \left . \frac{\partial \Delta \Psi (r,\vartheta,\varphi)}{\partial r}\right |_{r\ll 1} \right \rangle / \frac{\partial  \Psi_{\rm ss} (r)}{\partial r}  =   - \frac{2}{3} \frac{ (M_1+M_2)}{M_1}  r^3 \equiv - \frac{2}{3} \frac{ (M_1+M_2)}{M_1}  \frac{R^3}{a^3} \ .
    \label{eq:dev_close_tocenter_1}
\end{eqnarray}

Converting to the unitless acceleration used throughout in this manuscript, and realising that $r$ used in here is equivalent to $r_1$, we have for the deviation in the average effective acceleration due to being in a binary:

\begin{equation}
    \Delta \eta=- \frac{2}{3}\frac{1+q}{q} \ r_1^3 \ \eta_{\rm ss} \ .
    \label{eq:deviation_close_tocenter}
\end{equation}

\noindent Here ${\eta_{\rm ss}}$ is the the average effective acceleration of a spherically symmetric single star, and  $q=M_1/M_2$. 

\subsection{Potential}

\label{app:potential}

In the limit $r\ll 1$, the second term that provides the effect due to having a companion is:

\begin{eqnarray}
    \Psi_{\rm c} (r,\vartheta,\varphi)&=&  -\frac{GM_2}{a} \left ( 1 + r\cos \varphi \sin \vartheta -\frac{r^2}{2}\right ) \ .
    \nonumber \\ 
\end{eqnarray}

After the integration over all angles, we have:

\begin{equation}
\left \langle \left . \Delta \Psi (r,\vartheta,\varphi)\right |_{r\ll 1} \right \rangle / \Psi_{\rm ss} (r)  =   \frac{3M_2^2 + 2 M_1 M_2}{2 M_1(M_1+M_2)} \ r + \frac{2M_1 + 5M_2}{6 M_1}  r^3  , 
\end{equation}

\noindent and, for the unitless potential, with $q=M_1/M_2$:

\begin{equation}
\Delta \xi =  \left ( \frac{3+2q}{2q(1+q)} r_1+\frac{2q+5}{6q}\  r_1^3 \right ) \xi_{\rm ss} \ .
 \label{eq:dev_xi}
\end{equation}

\subsection{Effective acceleration for non-point mass}

\label{app:eff_acc_np}

The derivation for none-point mass is similar to described in \ref{app:eff_acc}, with the difference that in the Equation \ref{eq:potential_app}, the total mass $M_1$ in the first term ($\Psi_{ss}$) is replaced by the local value of mass,

\begin{equation}
\Psi_{\rm ss,loc} (r,\vartheta,\varphi) =
    -\frac{GM_{\rm loc}}{a}\frac{1}{r} \ . 
\end{equation}
\noindent Here $M_{\rm loc}$ is the local mass coordinate inside the donor star at a distance $r$ from the star's center. Thus,  the effective acceleration of a single, non-point mass star at radius $r$ and with local mass $M_{\rm loc}$ is:

\begin{equation}
\frac{\partial  \Psi_{\rm ss,loc} (r)}{\partial r} =  \frac{GM_{\rm loc}}{ a} \frac{1}{r^2} .
    \label{eq:ss_nonpoint}
\end{equation}

\begin{eqnarray}
& & \left \langle \left . \frac{\partial \Delta \Psi (r,\vartheta,\varphi)}{\partial r}\right |_{r\ll 1} \right \rangle / \frac{\partial  \Psi_{\rm ss,loc} (r)}{\partial r}  =   - \frac{2}{3} \frac{ (M_1+M_2)}{M_{\rm loc}}  r^3 \equiv - \frac{2}{3} \frac{ (M_1+M_2)}{M_{\rm loc}}  \frac{R^3}{a^3} \ .
    \label{eq:dev_close_tocenter_non_point}
\label{eq:centr}
\end{eqnarray}

The unitless form of the following equation could be obtained if we use the fact that the scaling factor for effective acceleration \ref{eq:scalegrav} is the same for both terms of the numerator and the denominator of the left-hand side, we can write the left hand side as:
\begin{equation}
    \left \langle \left . \frac{\partial \Delta \Psi (r,\vartheta,\varphi)}{\partial r}\right |_{r\ll 1} \right \rangle / \frac{\partial  \Psi_{\rm ss,loc} (r)}{\partial r}  =    \frac{\Delta \eta_{\rm loc}}{\eta_{\rm ss,loc}} 
\label{eq:cnt_unitless}
\end{equation}

And, finally, using the definition of $q_{\rm loc}$:

\begin{equation}
    \Delta \eta_{\rm loc}=- \frac{2}{3}\frac{1+q}{q_{\rm loc}(r_1)} \ r_1^3 \ \eta_{\rm ss,loc} \ .
    \label{eq:deviation_close_tocenter_non_point}
\end{equation}

\bibliography{ref_rl}
\bibliographystyle{aasjournal}

\end{document}